\documentclass[aps,prb,showpacs,twocolumn,10pt,superscriptaddress,floatfix]{revtex4-2}
\usepackage[T1]{fontenc}
\usepackage[latin9]{inputenc}
\usepackage{soul}
\usepackage{ulem}
\usepackage{graphicx}
\usepackage{appendix}
\usepackage{amssymb}
\usepackage{amsmath}
\usepackage[abs]{overpic}
\usepackage{overpic}
\usepackage{bm}
\usepackage{overpic}
\usepackage[breaklinks=true,colorlinks,citecolor=blue,linkcolor=blue,urlcolor=blue]{hyperref}
\usepackage[bottom]{footmisc}
\usepackage{floatrow}
\usepackage{upgreek}
\usepackage[labelformat=parens,caption=false]{subfig}
\usepackage[usenames, dvipsnames]{color}

\usepackage{empheq}
\usepackage[most]{tcolorbox}

\allowdisplaybreaks

\begin{document}

\title{Plasmon-magnon interactions in two-dimensional honeycomb magnets}
\author{Sayandip Ghosh}
\affiliation{Istituto Italiano di Tecnologia, Graphene Labs, Via Morego 30, I-16163 Genova, Italy}
\affiliation{Dipartimento di Fisica dell'Universit\`a di Pisa, Largo Bruno Pontecorvo 3, I-56127 Pisa, Italy}
\affiliation{Department of Physics, Visvesvaraya National Institute of Technology Nagpur, 440010 Nagpur, India}
\author{Guido Menichetti}
\email{guido.menichetti@df.unipi.it}
\affiliation{Dipartimento di Fisica dell'Universit\`a di Pisa, Largo Bruno Pontecorvo 3, I-56127 Pisa, Italy}
\affiliation{Istituto Italiano di Tecnologia, Graphene Labs, Via Morego 30, I-16163 Genova, Italy}
\author{Mikhail I. Katsnelson}
\affiliation{Institute for Molecules and Materials, Radboud University, NL-6525 AJ Nijmegen, The Netherlands}
\author{Marco Polini}
\affiliation{Dipartimento di Fisica dell'Universit\`a di Pisa, Largo Bruno Pontecorvo 3, I-56127 Pisa, Italy}
\affiliation{Istituto Italiano di Tecnologia, Graphene Labs, Via Morego 30, I-16163 Genova, Italy}
\affiliation{ICFO-Institut de Ci\`{e}ncies Fot\`{o}niques, The Barcelona Institute of Science and Technology, Av. Carl Friedrich Gauss 3, 08860 Castelldefels (Barcelona),~Spain}
\date{\today}

\begin{abstract}
Two-dimensional honeycomb ferromagnets offer the unprecedented opportunity to study interactions between collective modes that in standard bulk ferromagnets do not cross paths. Indeed, they harbor an optical spin-wave branch, i.e. a spin wave which disperses weakly near the Brillouin zone center. When doped with free carriers, they also host the typical gapless plasmonic mode of 2D itinerant electron/hole systems. When the plasmon branch meets the optical spin-wave branch, energy and momentum matching occurs, paving the way for interactions between the charge and spin sector. In this Article we present a microscopic theory of such plasmon-magnon interactions, which is based on a double random phase approximation. We comment on the possibility to unveil this physics in recently isolated 2D honeycomb magnets such as ${\rm Cr}_2{\rm Ge}_2{\rm Te}_6$.
\end{abstract}

\maketitle
\section{Introduction}

Two-dimensional (2D) materials, first isolated in 2004, have been investigated for almost two decades~\cite{Novoselov_Science_2004,Novoselov_PNAS_2005,Novoselov_Nature_2005,Zhang_Nature_2005} and have been the source of truly spectacular discoveries~\cite{katsnelson,Geim2013}. Magnetic 2D materials, in particular, have been recently discovered~\cite{Gong2017,Huang2017} and are currently under a very active scrutiny~\cite{Burch2018,Gong2019,Gibertini2019,Soriano2020,Yang2020,Wei2021,Jiang2021,Santos2022}. However, despite these efforts, collective charge (i.e.~plasmons) and spin excitations ~(i.e.~magnons) in them remain relatively unexplored. Although investigations of spin resonance and magnetic spectroscopy in the presence of an external magnetic field have been recently carried out~\cite{Zhang2020,Cenker2020}, it is extremely difficult to investigate spin-wave excitations in 2D magnetic samples by conventional momentum-resolved probes like neutron scattering. 

\begin{figure}[h]
\includegraphics[width=0.99\columnwidth]{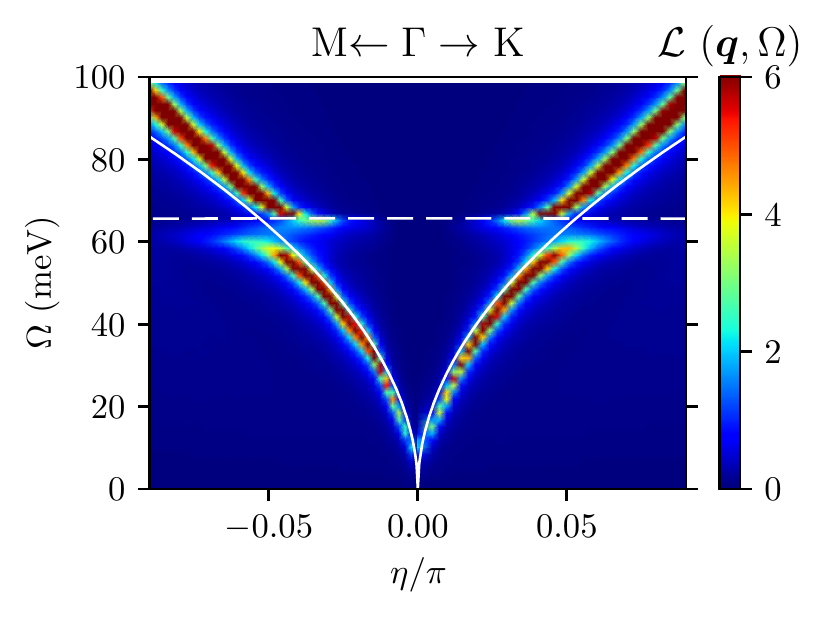}\\
\caption{(Color online) The calculated electron energy-loss function ${\cal L}({\bm q}, \Omega)$ (color scale)---defined in Eq.~(\ref{Eq:loss})---of monolayer doped Cr$_2$Ge$_2$Te$_6$ is plotted as a function of $\Omega$ and ${\bm q}$, around the $\Gamma$ point. These results have been obtained by including electron-magnon interactions, which manifest as plasmon-magnon interactions. The bare plasmon and bare magnon dispersions are shown by solid and dashed white lines, respectively. Here, we have taken a background dielectric constant $\kappa = 3.0$, temperature $T= 20~{\rm K}< T_{\rm C}$, and a $1 \%$ hole doping corresponding to a $\approx 2.5 \times 10^{12}~{\rm cm}^{-2}$ hole density. Finally, the parameter $\eta$ is used to indicate the wave vector ${\bm q}$ along high-symmetry directions. The unit of $\eta$ is 0.2537 {\AA}$^{-1}$.\label{fig:loss}}
\end{figure}

Finding unconventional ways of probing spin waves in atomically-thin magnetic crystals is therefore of fundamental importance. The idea behind this Article is extremely simple. If spin waves would couple to excitations in the charge channel, e.g.~plasmons, one could access them---although indirectly---by probing charge excitations via established experimental tools such as electron energy-loss spectroscopy (EELS)~\cite{Eberlein_PRB_2008,Liu_PRB_2008,Koch_PRB_2010,Senga_Nature_2019} and scattering near-field optical spectroscopy (SNOM)~\cite{Grigorenko_Naturephoton_2012,Basov_Science_2016,Low_NatureMater_2017,Basov_Nanophotonics_2021,Plantey2021}. 

The first studies of the interaction between electrons and spin waves in magnetic conductors were carried out many decades ago~\cite{Hertz_1973,Edwards_1973,Nolting_1979,Nagaev,Moriya,vonsovsky1974magnetism,Yosida_Magnetism,Katsnelson_RMP_2008}. However, the interaction between plasmon and magnon excitations received very little attention. The reason is obvious. In three-dimensional (3D) conductors, the plasmon energy scale ($1$-$20~{\rm eV}$)  is much larger than the typical energy scale of magnons ($10$ - $100~{\rm meV}$)~\cite{Grosso2014,Coey2012}. Early calculations by Baskaran and Sinha revealed a negligible renormalization of plasmon and magnon energies~\cite{Baskaran1973}. This was confirmed by Barna{\'s}~\cite{Barnas1979} who showed that even though a mechanism of plasmon-magnon coupling can be realized in the presence of an external electric or magnetic field, no measurable hybridization of plasmon and magnon modes can be obtained in 3D. The situation is expected to be dramatically different in 2D doped magnetic materials. Indeed, these must, on the one hand, harbor a gapless plasmon mode~\cite{Giuliani2005,Mahan2000} with a dispersion relation $\omega \propto \sqrt{q}$ in the charge sector (see white solid line in Fig.~\ref{fig:loss}). On the other hand, 2D magnetic materials with a honeycomb crystal structure support spin waves with two branches, one of which is a weakly dispersive optical branch~\cite{Owerre_JPCM_2016,Fransson_PRB_2016,Costa_2Dmater_2020,Chen_PRX_2018} (see dashed white line in Fig.~\ref{fig:loss}). Consequently, a 2D wave vector ${\bm q}$ must exist at which plasmons and magnons share similar energy scale, thereby providing us the missing ingredient necessary to realize and study plasmon-magnon interactions.

In this Article, we present a microscopic theory of plasmon-magnon interactions in 2D doped magnetic materials. More precisely, by including electron-electron and electron-magnon interactions in the realm of a ``double random phase approximation'', we calculate the electron-energy loss function of a prototypical 2D honeycomb magnet, i.e.~monolayer doped ${\rm Cr}_2{\rm Ge}_2{\rm Te}_6$ (see Ref.~\onlinecite{Menichetti2019} and references therein to earlier work). We find that plasmon-magnon interactions clearly manifest in the spectra of collective charge excitations, yielding a splitting of the plasmon dispersion into two branches as shown in Fig.~\ref{fig:loss}.  This splitting can be probed by cryogenic SNOM~\cite{Yang2013,Ni2018,Chen2019,Sun2020} opening up a new avenue to investigate the properties of spin waves in 2D materials. Finally, we emphasize that, in stark contrast to common wisdom~\cite{Barnas1979,Efimkin2021,Dyrdal2022}, plasmon-magnon interactions do not require spin-orbit coupling to exist. We show that they naturally arise from the exchange interaction between the itinerant carriers and the localized magnetic moments hosted by the lattice. 

Our Article is organized as follows. In Sec.~\ref{sect:effective_Hamiltonian} we present the microscopic Hamiltonian we have used  to capture the physics we are interested in.  Sec.~\ref{sect:der_eff_ham} contains a derivation of the {\it effective} Hamiltonian, obtained by using a canonical transformation approach, which is carried out after introducing the eigenstate (band) representation for the itinerant electron subsystem and the magnon representation for the localized electron subsystem. In Sec.~\ref{sect:EELS} we describe the double random phase approximation (RPA) we have used  to evaluate the electron-energy loss function and present numerical results obtained by using it.
Finally, in Sec.~\ref{sect:conclusions} we present a brief summary of our main results. Two appendices contain further theoretical and numerical details. In Appendix~\ref{SM_eq_of_motion} we demonstrate how the exact same results can be obtained by an equation-of-motion (rather than a diagrammatic) approach. In Appendix~\ref{SM_temp} we discuss the role of temperature on the electron-energy loss spectra.

\section{Effective Hamiltonian}
\label{sect:effective_Hamiltonian}

In this Article we are interested in the possibility of coupling the collective modes (spin waves) of a 2D honeycomb magnet with the plasmons of an itinerant electron gas. To this end, we consider, as a prototypical example of a 2D honeycomb magnet~\cite{Gibertini2019},  monolayer Cr$_2$Ge$_2$Te$_6$ (MCGT).

\begin{figure}[t]
\begin{overpic}[width=0.9\textwidth]{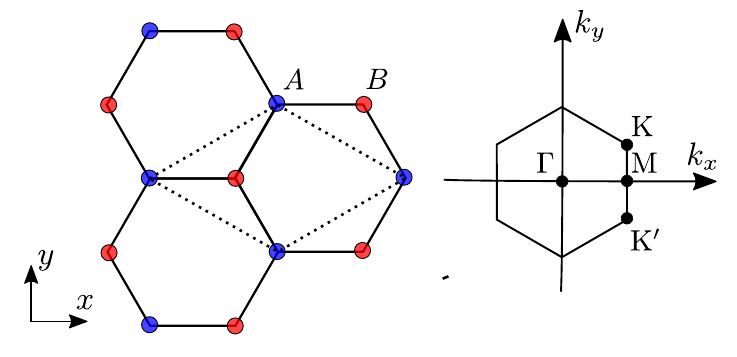} 
\put(1,95){\rm{(a)}} 
\put(130,95){\rm{(b)}}
\end{overpic}
\caption{(Color online) (a)~The honeycomb lattice formed by ${\rm Cr}$ atoms. $A$ and $B$ indicate the two sublattices. In panel (b) we show the corresponding Brillouin zone with its high symmetry points.
\label{fig:honeycomb}}
\end{figure}

MCGT is a crystal with rhombohedral $R\bar{3}$ symmetry formed by edge-sharing distorted CrTe$_6$ octahedra~\cite{Carteaux1995,Siberchicot1996}.
In this crystal, the ${\rm Cr}$ atoms form a honeycomb lattice which is  shown in Fig.~\ref{fig:honeycomb}(a). The unit cell (depicted by a dotted line) consists of two inequivalent atoms. $A$ and $B$ indicate the sublattice indices. The corresponding Brillouin zone is hexagonal. Its high symmetry points are shown in Fig.~\ref{fig:honeycomb}(b): $\Gamma = \left( 0,0 \right)$, $K = \left(2\pi/3, 2\pi/(3\sqrt{3})\right)$, $M = \left(2\pi/3,0 \right)$ and $K^\prime = \left(2\pi/3,- 2\pi/(3\sqrt{3})\right)$. Here, the ${\rm Cr}$-${\rm Cr}$ distance is taken to be unity for convenience. 

 According to {\it ab-initio} calculations (see Ref.~\onlinecite{Menichetti2019} and references therein), MCGT is an easy-axis ferromagnetic semiconductor with a gap $E_{\rm g} \approx 0.69~{\rm eV}$ and a Curie temperature $T_{\rm C} \approx 34~{\rm K}$~\cite{footnote_Tc}. Upon e.g.~electrostatic doping, free electrons/holes roam into this crystal and, at temperatures $T<T_{\rm C}$, co-exist with the ordered localized magnetic moments. For the sake of analytical amenability, we consider a model with two electronic subsystems: a) an itinerant 2D electron gas describing the aforementioned carriers due to doping and b) a localized electron system, contributing to the magnetic moments. Here, we are interested in the small doping regime, in which the Cr magnetic moment is well defined and independent of carrier concentration~\cite{Menichetti2021}. Subsystem a) alone displays 2D plasmons. Subsystem b) alone displays spin waves.  

The simplest Hamiltonian $\hat{\cal H}$ that captures the physics we are interested in contains four terms, i.e.~
\begin{equation}
\hat{\cal H} = \hat{\cal H}_{\rm s} + \hat{\cal H}_{\rm e} + \hat{\cal H}_{\rm es} + \hat{\cal H}_{\rm ee}.
\end{equation}
In our model, the localized magnetic moments are described by a ferromagnetic Heisenberg~\cite{Gubanov1992,Auerbach1994} $S=3/2$ Hamiltonian on a honeycomb lattice:
\begin{equation}\label{eq:Heisenberg+MAE}
\hat{\cal H}_{\rm s} = - \sum_{i<j} J_{ij} \hat{{\bm S}}_i \cdot \hat{{\bm S}}_j - \tilde{A} \sum_i \big(\hat{S}_{i}^{z}\big)^2~,
\end{equation} 
where $J_{ij}>0$ are the exchange couplings and $\tilde{A}$ is the magneto-crystalline anisotropy energy (MAE). 
We include first-, second-, and third-neighbor exchange processes, parametrized by three exchange constants, $J_1$, $J_2$, and $J_3$, respectively~\cite{Menichetti2019}. The MAE term circumvents the Mermin-Wagner theorem~\cite{Mermin1966,footnote_DM} and is responsible for long-range magnetic order in 2D~\cite{Irkhin1999}. 
The values of the exchange couplings and MAE used here have been calculated in Ref.~\cite{Menichetti2019} (see the caption of Fig.~\ref{fig:band_sw}). 

The second term in $\hat{\cal H}$, i.e.~$\hat{\cal H}_{\rm e}$, describes itinerant carriers. In this Article we choose a minimal one-orbital model, which captures the low-energy physics of MCGT (rather than developing a full tight-binding model to fit all the bands obtained from the {\it ab initio} calculations~\cite{Menichetti2019}). The electronic Hamiltonian in the site representation is given by
\begin{align}\label{eq:electronic_Hamilonian}
\hat{\cal H}_{\rm e} = & \sum_{i, j, \alpha} t_{ij} \hat{c}_{i,\alpha}^\dagger \hat{c}_{j,\alpha} + \lambda_{\rm SO} \sum_{\langle i,j \rangle, \alpha,\beta} \hat{c}_{i,\alpha}^\dagger \sigma^z_{\alpha \beta}  \hat{c}_{j,\beta} \nonumber\\
& +  U \sum_{i, \alpha} \delta_i \hat{c}_{i,\alpha}^\dagger \hat{c}_{i,\alpha}~,
\end{align} 
where Latin indices $i$ and $j$ denote the sites of a 2D honeycomb lattice (which is not a Bravais lattice and therefore a sublattice index $\tau$ will appear momentarily). In Eq.~(\ref{eq:electronic_Hamilonian}), $\hat{c}_{i,\alpha}^\dagger$ ($\hat{c}_{j,\beta}$) creates (annihilates) an electron with spin $\alpha=\uparrow,\downarrow$ ($\beta$) on site $i$ ($j$),  $t_{ij}$-which physically represents the tunneling amplitude between sites $i$ and $j$---is a real, symmetric matrix that limits hopping processes to first-, second- and third-neighbor sites. The effect of spin-orbit coupling is accounted for by an effective spin-dependent nearest-neighbor hopping of strength $\lambda_{\rm SO}$~\cite{Kane2005,Shitade2009,Foyevtsova2013,Mohapatra2019}. This term is present since MCGT does not have a mirror symmetry about the $xy$ plane. Once again, this term does not play any {\it qualitative} role and plasmon-magnon interactions occur even for $\lambda_{\rm SO} = 0$ (and in fact weakly depend on $\lambda_{\rm SO} \neq 0$). Finally, the third term in Eq.~(\ref{eq:electronic_Hamilonian}) is a staggered---$\delta_i = \pm 1$---sublattice potential of amplitude $U$, which is present since MCGT is not invariant under a $180^{\circ}$ rotation around the $z$ axis.

The third term in the full Hamiltonian $\hat{\cal H}$, i.e.~$\hat{\cal H}_{\rm es}$, describes the interaction between the itinerant electron's spin and the localized magnetic moments:
\begin{equation}\label{electron-spin}
\hat{\cal H}_{\rm es} = -{\cal I} \sum_{i,\alpha,\beta} \hat{{\bm S}}_i \cdot \left(\hat{c}_{i,\alpha}^\dagger {\bm \sigma}_{\alpha \beta} \hat{c}_{i,\beta}\right)~,
\end{equation}
where ${\cal I}>0$ is a ferromagnetic on-site exchange coupling between
localized and itinerant electrons, not to be mixed with Heisenberg
exchange interactions.

Finally, the last term in $\hat{\cal H}$ describes long-range Coulomb interactions between the itinerant electrons:
\begin{equation}\label{eq:Coulomb_general}
\hat{\cal H}_{\rm ee}= \frac{1}{2}\sum_{i \neq j,\alpha,\beta}\dfrac{e^2}{\kappa|{\bm r}_i - {\bm r}_j|} \hat{c}_{i,\alpha}^\dagger \hat{c}_{j,\beta}^\dagger \hat{c}_{j,\beta} \hat{c}_{i,\alpha}~,
\end{equation} 
where $\kappa$ is a background dielectric constant. Extrinsic effects due to nearby hyperbolic media~\cite{woessner_naturemat_2015} and metal gates~\cite{Alonso_naturenano_2017,lundeberg_science_2017}, which lead to an effective momentum- and frequency-dependent dielectric screening function, can be easily taken into account and will be the subject of a future technical publication. The Coulomb term in Eq.~(\ref{eq:Coulomb_general}) is responsible for the occurrence of plasmons in the itinerant electron system~\cite{Giuliani2005}. Rather than ``bosonizing'' the latter system by resorting to a plasmon-pole approximation~\cite{Overhauser1971}, we here derive plasmons microscopically from $\hat{\cal H}_{\rm ee}$. 

\section{Derivation of the final effective Hamiltonian}
\label{sect:der_eff_ham}

We now move on to analyze the collective modes of the full Hamiltonian $\hat{\cal H}$. Before doing that, we first  briefly review the theory of spin waves in a 2D honeycomb magnet and then perform a change of basis in the electronic Hamiltonian, introducing the bare electronic bands (i.e.~bands in the absence of electron-electron interactions and coupling to the localized magnetic moments).

\begin{figure}[tp]
\centering
\begin{tabular}{c}
\begin{overpic}[width=0.8\columnwidth]{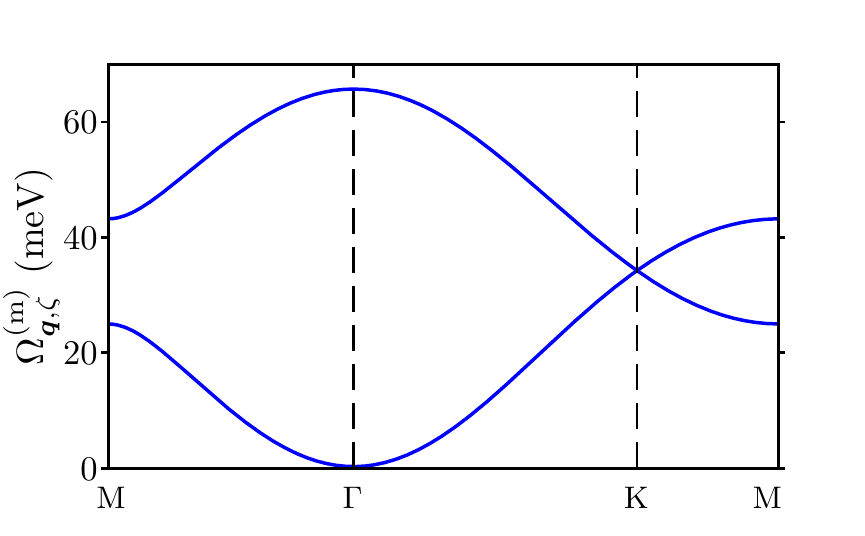} 
\put(0,115){\rm{(a)}}
\end{overpic} \\
\begin{overpic}[width=0.8\columnwidth]{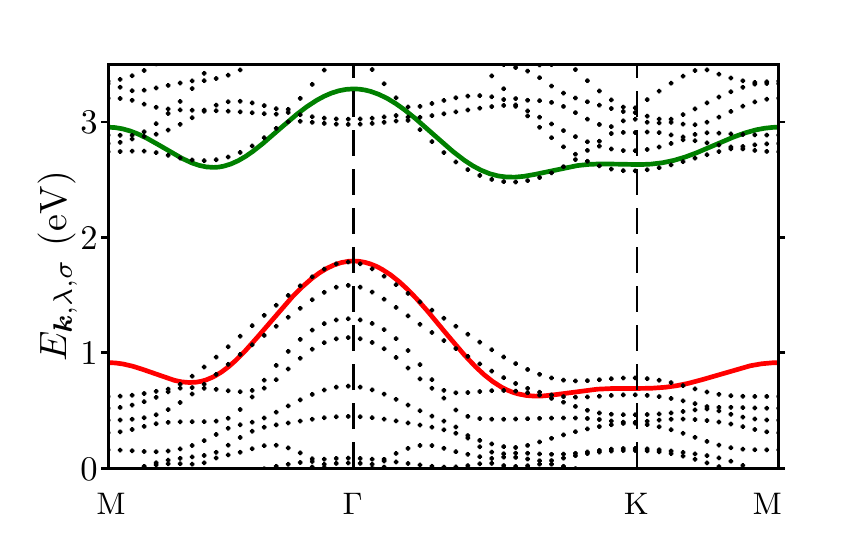}
\put(0,115){\rm{(b)}} 
\end{overpic}
\end{tabular}
\caption{(Color online) (a) Magnon dispersion for the following Heisenberg-model parameters $J_1 = 6.236~{\rm meV}$, $J_2 = 0.083~{\rm meV}$, $J_3= 0.268~{\rm  meV}$, and $\tilde{A} = 80~{\rm \mu eV}$---see Ref.~\onlinecite{Menichetti2019}. The lower spin-wave branch is gapped at $\Gamma$ but the gap is not visible on the scale of the figure, since it is given by $3S\tilde{A}\sim 0.4~{\rm  meV}$. (b) The red and green curves are the valence and conduction bands $E_{{\bm k}, \lambda, \sigma}$ of our non-interacting itinerant electron model described by Eq.~(\ref{eq:electronic_Hamilonian}). The black points are the results obtained from {\it ab initio} calculations~\cite{Menichetti2019}. The parameters of our model (\ref{eq:electronic_Hamilonian}) are: $t_1 = 0.32~{\rm eV}$, $t_2 = -0.04~{\rm eV}$, $t_3 = 0.45~{\rm eV}$, $\lambda_{\rm SO} = 0.09~{\rm eV}$, $U= 1.54~{\rm eV}$, and the band splitting $M \equiv {\cal I}S = 0.97~ \rm eV$.\label{fig:band_sw}}
\end{figure}

Below $T_{\rm C}$, the system of magnetic moments displays spin waves. These are found by following a standard linear-spin-wave analysis~\cite{Auerbach1994}, which we now briefly review. Such theory, which does not include quantum fluctuations, is valid for $T \ll T_{\rm C}$. Quantum effects in 2D magnets have been studied in the self-consistent spin-wave theory and have been shown to lead to small quantitative differences in the explored temperature range~\cite{Irkhin1999}.

We first write the spin operators $\hat{\bm S}_i$ in a Holstein-Primakoff representation~\cite{Auerbach1994,Holstein1904}. Retaining terms {\it up to} second order in the magnon operators, we have:
\begin{equation}
\label{eq:magnon}
\hat{S}_{i}^{z} \approx S - \hat{\xi}_{i}^\dagger \hat{\xi}_{i}~, ~ \hat{S}_{i}^{+} \approx \hat{\xi}_{i} \sqrt{2S}~, ~ \hat{S}_{i}^{-} \approx \hat{\xi}_{i}^\dagger \sqrt{2S}~.
\end{equation}
Here, $ \hat{\xi}_{i}^\dagger ( \hat{\xi}_{i})$ refers to the magnon creation (annihilation) operator for the $i$-th site.  Since we are dealing with a 2D honeycomb lattice with two atoms per unit cell, from here onward we change from the notation where ``$i$'' is the site index to the notation with a pair of indices, ``$i,\tau$'', where now $i$ indicates the unit-cell index and $\tau= A, B$ the sublattice index. The corresponding operators in momentum space, i.e.~$ \hat{\xi}_{{\bm q},\tau}^\dagger$ and $\hat{\xi}_{{\bm q},\tau}$, can be easily obtained by Fourier transforming $\hat{\xi}_{i,\tau}^\dagger $ and $\hat{\xi}_{i,\tau}$, respectively:
\begin{equation}
\hat{\xi}^\dagger_{i,\tau} = \frac{1}{\sqrt{N}}\sum_{{\bm q}} e^{ i {\bm q} \cdot {\bm R}_{i}}\hat{\xi}^\dagger_{{\bm q},\tau}
\end{equation}
and
\begin{equation}
\hat{\xi}_{i,\tau} = \frac{1}{\sqrt{N}}\sum_{{\bm q}} e^{- i {\bm q} \cdot {\bm R}_i}\hat{\xi}_{{\bm q},\tau}~.
\end{equation}
Here, $N$ is the number of unit cells, $\bm{q}$ belongs to the first Brillouin zone, and ${\bm R}_i$ denotes the position of the $i$-th unit cell.

After some algebraic calculation, the spin Hamiltonian (\ref{eq:Heisenberg+MAE}) can be written in terms of magnon operators as following:
\begin{equation}
\hat{\cal H}_{\rm s} = \hat{\cal H}^\prime_{\rm s} + {\rm constant}
\end{equation}
where
\begin{align}
\hat{\cal H}^\prime_{\rm s} &= \sum_{\bm q} \hat{\Phi}_{\bm q}^\dagger \left( \begin{array}{c c}
{\cal A}_{\bm q} & {\cal B}_{\bm q}  \\
{\cal B}^*_{\bm q} & {\cal A}_{\bm q} \end{array} \right) \hat{\Phi}_{\bm q} 
\end{align}
with $\hat{\Phi}_{\bm q} = (\hat{\xi}_{{\bm q},A} \quad \hat{\xi}_{{\bm q},B})^T$ and we have defined
\begin{align}
{\cal A}_{\bm q} =&S \Bigg[ 3J_1 + 3J_3 + 2 A \nonumber \\ & + 2 J_2 \Big(3 - \cos \sqrt{3}q_y - 2 \cos \frac{3 q_x}{2} \cos \frac{\sqrt{3} q_y}{2}\Big) \Bigg]~, 
\end{align}
and
\begin{align}
{\cal B}_{\bm q} =&S \Bigg[- J_1 \Big(e^{i q_x} + 2 e^{-i q_x/2} \cos \frac{\sqrt{3} q_y}{2}\Big ) \nonumber \\ &  - J_3 \Big (e^{- 2i q_x} + 2 e^{i q_x} \cos \sqrt{3} q_y\Big)\Bigg]~.
\end{align}

We now carry out a change of basis in order to diagonalize the magnon Hamiltonian:
\begin{equation}
\hat{\xi}_{{\bm q},\tau} = \sum_\zeta {\Lambda}_{\tau,\zeta}({\bm q}) \hat{\mu}_{{\bm q},\zeta}~, \qquad \hat{\xi}^\dagger_{{\bm q},\tau}  = \sum_\zeta {\Lambda}^\dagger_{\tau,\zeta}({\bm q}) \hat{\mu}_{{\bm q},\zeta}^\dagger~.
\end{equation}
The matrix form of the unitary transformation ${\Lambda}_{\tau, \zeta}$ reads as following:
\begin{equation}
{\Lambda} = \frac{1}{\sqrt{2}} \left( \begin{array}{c c}
e^{i \psi_{\bm q}/2} & e^{i \psi_{\bm q}/2} \\
e^{- i \psi_{\bm q}/2} & -e^{- i \psi_{\bm q}/2} 
\end{array}\right)~,
\end{equation}
where $\psi_{\bm q} = \arctan[{\rm Im}({\cal B}_{\bm q})/{\rm Re}({\cal B}_{\bm q})]$ and $\zeta= \pm 1$ refers to the magnon bands with energy $\Omega^{\rm (m)}_{{\bm q},\zeta}$ given by
\begin{equation}
\Omega^{\rm (m)}_{{\bm q},\pm} = {\cal A}_{\bm q} \pm |{\cal B}_{\bm q}|~.
\end{equation}

In such a linearized spin-wave approximation, the Heisenberg Hamiltonian is diagonal and reads as following:
\begin{equation}\label{eq:Heisenberg-spin-wave}
\hat{\cal H}^\prime_{\rm s} = \sum_{{\bm q}, \zeta} \Omega^{\rm (m)}_{{\bm q},\zeta} \hat{\mu}_{{\bm q},\zeta}^\dagger \hat{\mu}_{{\bm q},\zeta}~,
\end{equation}
where $\Omega^{\rm (m)}_{{\bm q},\zeta}$ is the magnon energy for wave vector ${\bm q}$ and magnon-band index $\zeta = \pm 1$. In writing Eq.~(\ref{eq:Heisenberg-spin-wave}) we have neglected a trivial constant. The two magnon branches, $\Omega^{\rm (m)}_{{\bm q}, \zeta}$ for $\zeta = \pm 1$, are plotted as functions of ${\bm q}$ in Fig.~\ref{fig:band_sw}(a). While the energy of the lower magnon branch is too small near the $\Gamma$ point (${\bm q} = {\bm 0}$), the energy of the upper branch is much larger, paving the way for the possibility to study plasmon-magnon interactions near the $\Gamma$ point. 

We now work on the electron-spin coupling Hamiltonian in Eq.~\eqref{electron-spin}. Performing the approximate Holstein-Primakoff transformation (\ref{eq:magnon}) for the spin operators, Eq.~\eqref{electron-spin} can be written in terms of magnon operators, in the momentum-space representation, as:
\begin{align}
\label{eq:espin}
\hat{\cal H}_{\rm es} =&-{\cal I} S \sum_{{\bm k},\tau,\sigma} \sigma \hat{c}_{{\bm k},\tau,\sigma}^\dagger \hat{c}_{{\bm k},\tau,\sigma}\nonumber  \\ &+ {\cal I} \sum_{{\bm k},\tau,\sigma} \sum_{{\bm q},{\bm p}} \sigma  \hat{\xi}_{{\bm q},\tau}^\dagger  \hat{\xi}_{{\bm q}+{\bm p},\tau} \hat{c}_{{\bm k},\tau,\sigma}^\dagger \hat{c}_{{\bm k}-{\bm p},\tau,\sigma} \nonumber \\
&-{\cal I} \sqrt{2S} \sum_{{\bm k},\tau} \sum_{\bm q} \left( \hat{\xi}^\dagger_{{\bm q},\tau} \hat{c}_{{\bm k},\tau,\uparrow}^\dagger \hat{c}_{{\bm k}+{\bm q},\tau,\downarrow} + \right.  \nonumber \\ &\left. \hat{\xi}_{{\bm q},\tau} \hat{c}_{{\bm k}+{\bm q},\tau,\downarrow}^\dagger \hat{c}_{{\bm k},\tau,\uparrow}\right)~,
\end{align}
where $\sigma = +(-)$ refers to spin $\uparrow$ ($\downarrow$) and $\hat{c}^{\dagger}_{{\bm k},\tau,\sigma}$, $\hat{c}_{{\bm k},\tau,\sigma}$ are the wave-vector (Bloch) representation of the electronic field operators:
\begin{equation}
\label{eq:cdagger}
\hat{c}^\dagger_{i,\tau,\sigma} = \frac{1}{\sqrt{N}}\sum_{\bm q} e^{i {\bm q} \cdot {\bm R}_i}\hat{c}^\dagger_{{\bm q},\tau,\sigma}
\end{equation}
and
\begin{equation}
\label{eq:c}
\hat{c}_{i,\tau,\sigma} = \frac{1}{\sqrt{N}}\sum_{\bm q} e^{- i {\bm q} \cdot {\bm R}_i}\hat{c}_{{\bm q},\tau,\sigma}~.
\end{equation}

The first term in Eq.~\eqref{eq:espin} corresponds to the polarization of the itinerant electrons by the ionic magnetization and will be included in the electronic Hamiltonian---see $\hat{\cal H}^\prime_{\rm e}$ in Eq.~(\ref{eq:electronic_Hamiltonian_con_un_pezzoinpiu}) below. The second and third terms represent electron-magnon interactions.

It is convenient to switch to a diagonal, band representation also in the case of the itinerant electron system. Inserting Eqs.~\eqref{eq:cdagger}-\eqref{eq:c} in Eq.~(\ref{eq:electronic_Hamilonian}) and taking into account the first term in Eq.~(\ref{eq:espin}), we finally find 
\begin{align}
\label{eq:electronic_Hamiltonian_con_un_pezzoinpiu}
\hat{\cal H}^\prime_{\rm e} &\equiv \hat{\cal H}_{\rm e} - M \sum_{{\bm k},\tau,\sigma} \sigma \hat{c}_{{\bm k},\tau,\sigma}^\dagger \hat{c}_{{\bm k},\tau,\sigma} = \nonumber\\ 
&= \sum_{{\bm k},\sigma} \left\{ \left[ \epsilon_2 ({\bm k}) - M \sigma + U \right] \hat{c}_{{\bm k},A,\sigma}^\dagger \hat{c}_{{\bm k},A,\sigma}\right. \nonumber \\ &\left. + \left[ \epsilon_2 ({\bm k}) - M \sigma - U\right] \hat{c}_{{\bm k},B,\sigma}^\dagger \hat{c}_{{\bm k},B,\sigma}\right. \nonumber \\ 
&\left. + \left[\epsilon_1 ({\bm k}) + \epsilon_3 ({\bm k}) + \sigma \epsilon_{\rm SO} ({\bm k})\right] \hat{c}_{{\bm k},A,\sigma}^\dagger \hat{c}_{{\bm k},B,\sigma}\right. \nonumber \\ &\left. + \left[\epsilon_1^* ({\bm k}) + \epsilon_3^* ({\bm k}) + \sigma \epsilon_{\rm SO}^* ({\bm k})\right]\hat{c}_{{\bm k},B,\sigma}^\dagger \hat{c}_{{\bm k},A,\sigma} \right\}~,
\end{align}
where
\begin{align}
& M  \equiv {\cal I} S~,\\
&\epsilon_1 ({\bm k}) = t_1 \left( e^{i k_x} + 2 e^{-i k_x/2} \cos \frac{\sqrt{3}k_y}{2} \right)~, \\
&\epsilon_2 ({\bm k}) = 2 t_2 \left( \cos \sqrt{3}k_y + 2 \cos \frac{3k_x}{2} \cos \frac{\sqrt{3}k_y}{2} \right)~, \\
&\epsilon_3 ({\bm k}) = t_3 \left( e^{- 2i k_x} + 2 e^{i k_x} \cos \sqrt{3}k_y \right)~,
\end{align}
and
\begin{align}
&\epsilon_{\rm SO} ({\bm k}) = \lambda_{\rm SO} \left( e^{i k_x} + 2 e^{-i k_x/2} \cos \frac{\sqrt{3}k_y}{2}\right)~.
\end{align} 

Due to the absence of a spin-mixing term, the tight-binding Hamiltonian (\ref{eq:electronic_Hamiltonian_con_un_pezzoinpiu}) in momentum space can be decoupled for $\uparrow$ and $\downarrow$ spins, and can be expressed in the Bloch spinor representation as
\begin{align}
\hat{\cal H}^\prime_{\rm e} &= \sum_{\bm k} \sum_\sigma \hat{\cal H}^\prime_{\rm e} ({\bm k},\sigma)
\end{align}
where
\begin{widetext}
\begin{align}
\hat{\cal H}^\prime_{\rm e} ({\bm k},\sigma) &= \hat{\Psi}_{{\bm k},\sigma}^\dagger \left( \begin{array}{c c}
\epsilon_2 ({\bm k}) - M \sigma + U & \epsilon_1 ({\bm k}) + \epsilon_3 ({\bm k}) + \sigma \epsilon_{\rm SO} ({\bm k})  \\
\epsilon_1^* ({\bm k}) + \epsilon_3^* ({\bm k}) + \sigma \epsilon_{\rm SO}^* ({\bm k}) & \epsilon_2 ({\bm k}) - M \sigma - U \end{array}\right)\hat{\Psi}_{{\bm k},\sigma}~,
\end{align}
\end{widetext}
and $\hat{\Psi}_{{\bm k},\sigma} = ( \hat{c}_{{\bm k},A,\sigma} \quad \hat{c}_{{\bm k},B,\sigma})^T$. 
It is more convenient at this stage to perform a change of basis, 
\begin{align}
\label{Eq:unitary_electron}
&\hat{c}_{{\bm k},\tau,\sigma} = \sum_\lambda \Gamma_{\tau,\lambda,\sigma} ({\bm k}) \hat{f}_{{\bm k},\lambda,\sigma}\nonumber\\ &\hat{c}^{\dagger}_{{\bm k},\tau,\sigma} = \sum_\lambda \Gamma^{\dagger}_{\tau,\lambda,\sigma} ({\bm k}) \hat{f}^{\dagger}_{{\bm k},\lambda,\sigma}~,
\end{align}
$\Gamma_{\tau,\lambda,\sigma}$ being a unitary transformation written in a matrix form as
\begin{equation}
\Gamma_{\sigma} = \left( \begin{array}{c c}
\alpha_{{\bm k},\sigma} e^{i \varphi_{{\bm k},\sigma}/2} & \beta_{{\bm k},\sigma} e^{i \varphi_{{\bm k},\sigma}/2} \\
\beta_{{\bm k},\sigma} e^{-i \varphi_{{\bm k},\sigma}/2} & -\alpha_{{\bm k},\sigma} e^{-i \varphi_{{\bm k},\sigma}/2} 
\end{array}\right)~.
\end{equation}
Here, we have defined
\begin{align}
\alpha_{{\bm k},\sigma} &= \tfrac{1}{\sqrt{2}} \left[ 1 + \tfrac{U}{\sqrt{|\epsilon_1({\bm k}) + \epsilon_3({\bm k}) + \sigma \epsilon_{\rm SO}({\bm k})|^2 + U^2}} \right]^\frac{1}{2}\nonumber \\
\beta_{{\bm k},\sigma} &= \tfrac{1}{\sqrt{2}} \left[ 1 - \tfrac{U}{\sqrt{|\epsilon_1({\bm k}) + \epsilon_3({\bm k}) + \sigma \epsilon_{\rm SO}({\bm k})|^2 + U^2}} \right]^\frac{1}{2}~,
\end{align}
and
\begin{equation}
\varphi_{{\bm k},\sigma} = \arctan \left[\frac{{\rm Im}(\epsilon_1 + \epsilon_3 + \sigma \epsilon_{\rm SO})}{{\rm Re}(\epsilon_1 + \epsilon_3 + \sigma \epsilon_{\rm SO})}\right]~.
\end{equation}
In Eq.~(\ref{Eq:unitary_electron}), $\hat{f}^{\dagger}_{{\bm k},\lambda,\sigma}$ ($\hat{f}_{{\bm k},\lambda,\sigma}$) creates  (destroys) an electron with wave vector $\bm k$,  band index $\lambda = \pm 1$, spin $\sigma = \uparrow, \downarrow$ and energy given by $E_{{\bm k},{\pm},\sigma}$

\begin{align}
E_{{\bm k},{\pm},\sigma} &= \epsilon_2({\bm k})  - M \sigma \nonumber \\ &\pm  \sqrt{|\epsilon_1({\bm k}) + \epsilon_3({\bm k}) + \sigma \epsilon_{\rm SO}({\bm  k})|^2 + U^2}~.
\end{align}

In this band basis, the electronic Hamiltonian $\hat{\cal H}_{\rm e}$ assumes the following diagonal form:
\begin{align}\label{eq:bare_fermionic_hamiltonian}
\hat{\cal H}^\prime_{\rm e}  =  \sum_{{\bm k},\lambda, \sigma} E_{{\bm k},\lambda,\sigma} \hat{f}^{\dagger}_{{\bm k},{\lambda},\sigma}\hat{f}_{{\bm k},{\lambda},\sigma}~.
\end{align}
The electronic band energies $E_{{\bm k}, \lambda, \sigma}$ are shown in Fig.~\ref{fig:band_sw}(b). These results have been obtained by fitting the results of accurate {\it ab initio} calculations~\cite{Menichetti2019} in the vicinity of the valence band maximum and conduction band minima. We stress that, for the sake of analytical amenability, in this work we have considered a minimal one-orbital model which describes the relevant low-energy physics of MCGT, instead of developing a full tight-binding model to mimic the {\it ab initio} band structure.

In the aforementioned semiclassical spin-wave $S\gg 1$ approximation and after expressing the longitudinal and spin-flip parts of the exchange Hamiltonian $\hat{\cal H}_{\rm es}$ in Eq.~(\ref{eq:electronic_Hamiltonian_con_un_pezzoinpiu}) in terms of the magnonic and electronic operators in the respective band representations, two terms are generated:
\begin{widetext}
\begin{align}
\hat{\cal H}_{1, {\rm em}} &= {\cal I} 
\sum_{{\bm k},{\bm q},{\bm p}} 
\sum_{\tau} \sum_{\lambda,\lambda^\prime} 
\sum_{\zeta, \zeta^\prime} 
\sum_{\sigma} \sigma  {\Lambda}^\dagger_{\tau, \zeta}({\bm q}) {\Lambda}_{\tau, \zeta^\prime}({\bm q}+{\bm p}) \Gamma_{\tau,\lambda,\sigma} ({\bm k}) \Gamma_{\tau,\lambda^\prime,\sigma} ({\bm k}-{\bm p}) \hat{\mu}_{{\bm q},\zeta}^\dagger \hat{\mu}_{{\bm q}+{\bm p},\zeta^\prime} \hat{f}^{\dagger}_{{\bm k},{\lambda},\sigma}\hat{f}_{{\bm k}-{\bm p},{\lambda^\prime},\sigma}~,\label{eq:H1em}\\
\hat{\cal H}_{2, {\rm em}} &= - {\cal I} \sqrt{2S} \sum_{{\bm k},{\bm q}} \sum_{\tau} \sum_{\lambda,\lambda^\prime} \sum_{\zeta} \left( {\Lambda}^\dagger_{\tau, \zeta}({\bm q}) \Gamma^\dagger_{\tau,\lambda,\uparrow} ({\bm k}) \Gamma_{\tau,\lambda^\prime,\downarrow} ({\bm k}+{\bm q}) \hat{\mu}_{{\bm q},\zeta}^\dagger \hat{f}^{\dagger}_{{\bm k},{\lambda},\uparrow}\hat{f}_{{\bm k}+{\bm q},{\lambda^\prime},\downarrow} + \rm{H.c.} \right)~. \label{eq:H2em}
\end{align} 
\end{widetext}
The term $\hat{\cal H}_{1, {\rm em}}$ describes direct scattering of magnons and electrons with strength ${\cal I} \sigma$ while the term $\hat{\cal H}_{2, {\rm em}}$ describes the emission or absorption of magnons with strength $-{\cal I}\sqrt{2S}$. 

Since states at the Fermi energy have only one spin~\cite{Menichetti2019}, $\hat{\cal H}_{2, {\rm em}}$, which is a spin-flip term, can be removed by a canonical Schrieffer-Wolff transformation~\cite{Nagaev1974,Katsnelson_RMP_2008} whose smallness parameter is $1/S\ll 1$. We note that $\hat{\cal H}_{2, {\rm em}}$ is {\it linear} in the magnon operators and describes the emission or absorption of magnons with strength $-{\cal I}\sqrt{2S}$. The Schrieffer-Wolff  transformation is a canonical transformation of the following form:
\begin{equation}
\hat{\cal H}^\prime = e^{\hat{P}} \hat{\cal H} e^{-\hat{P}} = \hat{\cal H} + [\hat{P},\hat{\cal H}] + \frac{1}{2} \left[\hat{P}, [\hat{P},\hat{\cal H}] \right] +\dots~.
\label{eq_canonical}
\end{equation}
With the aim of eliminating $\hat{\cal H}_{2, {\rm em}}$, the generator $\hat{P}$ of the transformation is chosen in such a way to satisfy the following equation:
\begin{equation}
[\hat{P},\hat{\cal H}^\prime_{\rm e} + \hat{\cal H}^\prime_{\rm s}] + \hat{\cal H}_{2,{\rm em}} = 0~.
\label{eq_canonical2}
\end{equation}
The general solution of the previous equation is: 
\begin{align}
\hat{P} =& - {\cal I} \sqrt{2S} \sum_{{\bm k},{\bm q},\tau,\lambda,\lambda^\prime,\zeta}  \left[ X_{\tau,\zeta,\lambda,\lambda^\prime}({\bm k},{\bm q}) {\Lambda}^\dagger_{\tau, \zeta}({\bm q}) \right. \nonumber \\ &\left. \times \Gamma^\dagger_{\tau,\lambda,\uparrow} ({\bm k}) \Gamma_{\tau,\lambda^\prime,\downarrow} ({\bm k}+{\bm q}) \hat{\mu}_{{\bm q},\zeta}^\dagger \hat{f}^{\dagger}_{{\bm k},{\lambda},\uparrow}\hat{f}_{{\bm k}+{\bm q},{\lambda^\prime},\downarrow}\right. \nonumber \\ 
&+ \left.Y_{\tau,\zeta,\lambda,\lambda^\prime}({\bm k},{\bm q}) {\Lambda}_{\tau, \zeta}({\bm q}) \Gamma^\dagger_{\tau,\lambda^\prime,\downarrow} ({\bm k}+{\bm q}) \Gamma_{\tau,\lambda,\uparrow} ({\bm k})\right. \nonumber \\ &\left. \times \hat{\mu}_{{\bm q},\zeta} \hat{f}^{\dagger}_{{\bm k}+{\bm q},{\lambda^\prime},\downarrow} \hat{f}_{{\bm k},{\lambda},\uparrow} \right]~,
\end{align}

where
\begin{align}
X_{\tau,\zeta,\lambda,\lambda^\prime}({\bm k},{\bm q})  &= - Y_{\tau,\zeta,\lambda,\lambda^\prime}({\bm k},{\bm q}) \nonumber  \\ &= - \frac{1}{E_{{\bm k}+{\bm q},\lambda^\prime,\downarrow} - E_{{\bm k},\lambda,\uparrow}- \Omega^{\rm (m)}_{{\bm q},\zeta}} \nonumber \\ &\approx - \frac{1}{h_{{\bm k}+{\bm q},\lambda^\prime} - h_{{\bm k},\lambda} + 2M}~.
\end{align}
We have defined $h_{{\bm k},\lambda} = \epsilon_2({\bm k})  + \lambda \sqrt{|\epsilon_1({\bm k}) + \epsilon_3({\bm k})|^2 + U^2}$. It is evident that $X_{\tau,\zeta,\lambda,\lambda^\prime}({\bm k},{\bm q})$ and $Y_{\tau,\zeta,\lambda,\lambda^\prime}({\bm k},{\bm q})$ do not depend on the sublattice index $\tau$ and on the magnon band index $\zeta$ anymore and we can write $X_{\tau,\zeta,\lambda,\lambda^\prime}=X_{\lambda,\lambda^\prime}$ and $Y_{\tau,\zeta,\lambda,\lambda^\prime}=Y_{\lambda,\lambda^\prime}$. 

The price that one has to pay for  removing $\hat{\cal H}_{2, {\rm em}}$ is a renormalization of $\hat{\cal H}_{1,\rm em}$, as we now show.  If we neglect the terms of order higher than quadratic in Eq.~(\ref{eq_canonical}), we obtain the following effective electron-magnon interaction:
\begin{equation}
\hat{\cal H}^\prime_{1, \rm em} = \hat{\cal H}_{1,\rm em} + \frac{1}{2}[P,\hat{\cal H}_{2
,\rm em}]~.
\end{equation}
After a tedious but straightforward calculation, we find
\begin{align}
\hat{\cal H}^\prime_{1,\rm em} =& \sum_{{\bm k},{\bm q},{\bm p}} \sum_{\tau,\sigma} \sum_{\lambda,\lambda^\prime,\zeta, \zeta^\prime} \sigma \tilde{\cal I}^{\lambda, \lambda^\prime}_{{\bm k},{\bm q},{\bm p}} {\Lambda}^\dagger_{\tau, \zeta}({\bm q}) {\Lambda}_{\tau, \zeta^\prime}({\bm q}+{\bm p}) \nonumber \\ &\times  \Gamma^\dagger_{\tau,\lambda,\sigma} ({\bm k}) \Gamma_{\tau,\lambda^\prime,\sigma} ({\bm k}-{\bm p}) \hat{\mu}_{{\bm q},\zeta}^\dagger \hat{\mu}_{{\bm q}+{\bm p},\zeta^\prime} \nonumber \\ &\times \hat{f}^{\dagger}_{{\bm k},{\lambda},\sigma}\hat{f}_{{\bm k}-{\bm p},{\lambda^\prime},\sigma} \nonumber \\
\equiv &\sum_{{\bm k},{\bm q},{\bm p}} \sum_{\tau, \sigma } \sum_{\lambda,\lambda^\prime, \zeta, \zeta^\prime}\sigma \tilde{\cal I}^{\lambda, \lambda^\prime}_{{\bm k},{\bm q},{\bm p}} A^{\zeta, \zeta^\prime; \lambda, \lambda^\prime}_{\tau, \sigma}({\bm k}, {\bm q}, {\bm p})\nonumber \\ &\times \hat{\mu}_{{\bm q},\zeta}^\dagger \hat{\mu}_{{\bm q}+{\bm p},\zeta^\prime} \hat{f}^{\dagger}_{{\bm k},{\lambda},\sigma}\hat{f}_{{\bm k}-{\bm p},{\lambda^\prime},\sigma}~,
\label{Eq:el-mag}
\end{align}
where the electron-magnon scattering amplitude $\tilde{\cal I}^{\lambda, \lambda^\prime}_{{\bm k},{\bm q},{\bm p}}$ is given by 
\begin{align}\label{eq:Jtilde}
\tilde{\cal I}^{\lambda, \lambda^\prime}_{{\bm k},{\bm q},{\bm p}} =& {\cal I} - {\cal I} M \left[ Y_{\lambda,\lambda^\prime}({\bm k},{\bm q}) + Y_{\lambda^\prime,\lambda}({\bm k}-{\bm p},{\bm q}+{\bm p}) \right]\nonumber \\
=& \frac{\cal I}{2} \left( \frac{h_{{\bm k}+{\bm q},\lambda^\prime} - h_{{\bm k},\lambda}}{h_{{\bm k}+{\bm q},\lambda^\prime} - h_{{\bm k},\lambda} + 2M} \right. \nonumber \\ &\left.  + \frac{h_{{\bm k}+{\bm q},\lambda} - h_{{\bm k}-{\bm p},\lambda^\prime}}{h_{{\bm k}+{\bm q},\lambda} - h_{{\bm k}-{\bm p},\lambda^\prime} + 2M} \right)~,
\end{align}
and we have introduced 
\begin{align}
A^{\zeta, \zeta^\prime; \lambda, \lambda^\prime}_{\tau, \sigma}({\bm k}, {\bm q}, {\bm p})\equiv &  {\Lambda}^\dagger_{\tau, \zeta}({\bm q}) {\Lambda}_{\tau, \zeta^\prime}({\bm q}+{\bm p}) \nonumber \\ &\times \Gamma^\dagger_{\tau,\lambda,\sigma} ({\bm k}) \Gamma_{\tau,\lambda^\prime,\sigma} ({\bm k}-{\bm p}).
\label{Eq:el-mag_A}
\end{align}
Note that $\tilde{\cal I}^{\lambda, \lambda^\prime}_{{\bm k},{\bm q},{\bm p}}$ vanishes for ${\bm q}, {\bm p} \to {\bm 0}$, thereby taking properly into account the rotational symmetry of electron-magnon interactions~\cite{Auslender1984,Irkhin_epjb_2002}.
A more general and rigorous variational approach, which is valid well beyond the regime of applicability of perturbation theory in $1/S$, does not produce qualitatively different outcomes~\cite{Auslender1983,Katsnelson_RMP_2008}. 

The final result for the transformed Hamiltonian is:
\begin{equation}\label{tot_hamil}
\hat{\cal H}^{\prime} = \hat{\cal H}^\prime_{\rm e} + \hat{\cal H}^\prime_{\rm s} + \hat{\cal H}^\prime_{1, {\rm em}} + \hat{\cal H}^\prime_{\rm ee}~, 
\end{equation}
where $\hat{\cal H}^\prime_{\rm e}$ is reported in Eq.~(\ref{eq:bare_fermionic_hamiltonian}), $\hat{\cal H}^\prime_{\rm s}$ in Eq.~(\ref{eq:Heisenberg-spin-wave}), and $\hat{\cal H}^\prime_{1, {\rm em}}$ in Eq.~\eqref{Eq:el-mag}.  The last term in Eq.~(\ref{tot_hamil}) is the electron-electron interaction Hamiltonian in the Fourier representation, $\hat{\cal H}_{\rm ee}= \frac{1}{2}\sum_{{\bm q} \neq {\bm 0}}v_{\bm q} \hat{\rho}({\bm q}) \hat{\rho}(-{\bm q})$, where $\hat{\rho} ({\bm q}) = \sum_{{\bm k},\tau,\sigma,\lambda,\lambda^\prime} \Gamma^\dagger_{\tau,\lambda,\sigma}({\bm k} + {\bm q})\Gamma_{\tau,\lambda^\prime,\sigma}({\bm k}) \hat{f}^{\dagger}_{{\bm k} + {\bm q},{\lambda},\sigma} \hat{f}_{{\bm k},{\lambda^\prime}\sigma}$ is the Fourier-transformed density operator and $v_{\bm q} = 2\pi e^2 / (\kappa q)$ is the 2D Fourier transform of the Coulomb potential~\cite{Giuliani2005}.

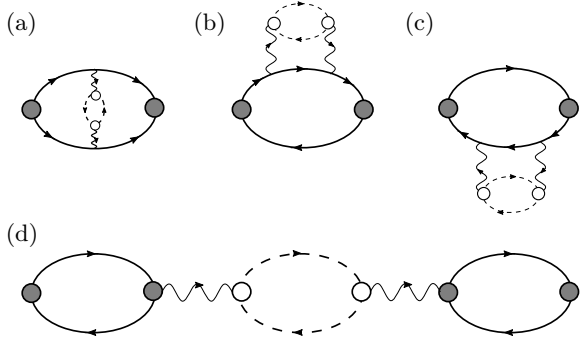
\begin{figure}[t]
\begin{overpic}[width=0.90\columnwidth]{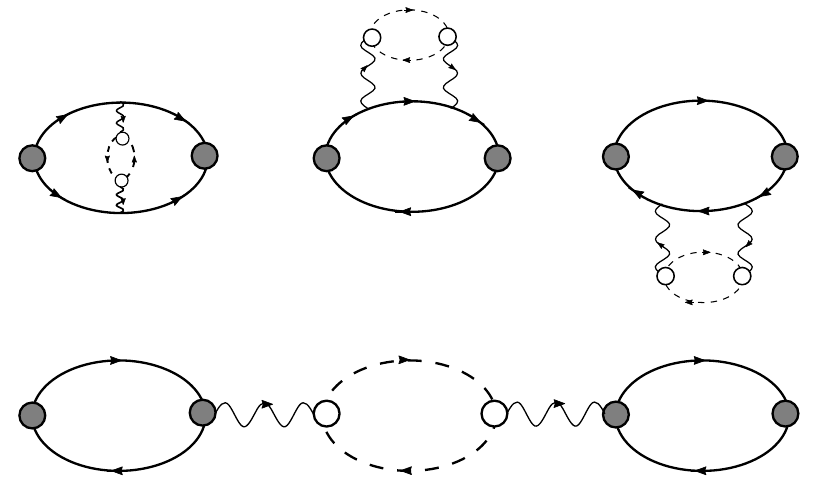} 
\put(-1,120){\rm{(a)}} 
\put(70,120){\rm{(b)}} 
\put(150,120){\rm{(c)}} 
\put(-1,40){\rm{(d)}}
\end{overpic}
\caption{Feynman diagrams for the spin-resolved density-density response function up to second order in the electron-magnon interaction. The solid [dashed] line refers to the free-electron $G^{(0)}_\sigma({\bm k}, i \omega_m)$ [free-magnon ${\cal G}^{(0)} ({\bm q}, i \Omega_n)$] Green's function; $i \omega_m = (2m +1)\pi/\beta$, $i \Omega_n = 2n\pi/\beta$ ($n,m = 0, 1, 2, 3, ...$) are the fermionic and bosonic Matsubara frequencies respectively with $\beta = (k_{\rm B}T)^{-1}$ being the inverse temperature. The wavy line corresponds to the electron-magnon scattering process. Finally, the filled grey vertices in the fermionic bubbles [the empty vertices in the magnonic bubbles] refer to the spin-resolved  electronic density operator $\hat{\rho}_{\sigma} ({\bm q})$ [magnon density operator $\hat{\rho}^{\rm (m)} ({\bm q}) = \sum_{{\bm k},\zeta} \hat{\mu}_{{\bm k}+{\bm q},\zeta}^\dagger \hat{\mu}_{{\bm k},\zeta}$]. }
\label{fig:second_order} 
\end{figure} 
\section{Theory of the electron-energy loss function}
\label{sect:EELS}
\subsection{Double RPA approach}

The spectrum of density excitations can be found, in principle, by finding the zeroes of the dynamical dielectric function $\varepsilon({\bm q}, \Omega)$ in the wave vector ${\bm q}$ and frequency $\Omega$ space~\cite{Giuliani2005}. Numerically, however, it is more convenient to calculate the so-called ``loss'' function ${\cal L} ({\bm q},\Omega)$, which can be directly measured in EELS~\cite{Egerton_Springer_2011,Westerhout2022}:
\begin{equation}
{\cal L} ({\bm q},\Omega) = - {\rm Im}\left[\frac{1}{\varepsilon({\bm q},\Omega)} \right]~,
\label{Eq:loss}
\end{equation}
where $\varepsilon({\bm q}, \Omega)$ is related to the causal density-density response function $\chi_{\rho \rho}  ({\bm q}, \Omega)$ by the usual relation~\cite{Giuliani2005}
\begin{equation}\label{eq:relation_epsilon_chi}
\frac{1}{\varepsilon({\bm q}, \Omega)} = 1+ v_{\bm q} \chi_{\rho\rho}({\bm q}, \Omega)~.
\end{equation}
To understand the interplay between electron-magnon and electron-electron interactions in the loss function, we calculate $\chi_{\rho\rho}({\bm q}, \Omega)$ by carrying  out {\it two} RPAs~\cite{Giuliani2005} in a row, the first one with respect to electron-magnon interactions and the second one with respect to electron-electron interactions. Below, we will refer to this procedure as to ``double RPA approach''. The exact same results, obtained here with a diagrammatic approach, can be obtained via an equation-of-motion approach~\cite{Mahan2000,Reijnders2022}, as detailed in Appendix.~\ref{SM_eq_of_motion}. 

In what follows we illustrate the steps that are necessary to calculate the density-density response function $\chi_{\rho \rho}  ({\bm q},i\Omega_n)$ in the imaginary-frequency domain, where diagrammatic perturbation theory is most conveniently formulated. 
The causal quantity $\chi_{\rho \rho}  ({\bm q},\Omega)$ is obtained after carrying out the usual analytical continuation $i\Omega_n \to \Omega + i 0^+$.

Our perturbative scheme relies on the knowledge of the non-interacting spin-resolved density-density response function, $\chi^{(0)}_{\rho \rho, \sigma} ({\bm q},i \Omega_n)$, which does not include any of the interactions at play. It is simply obtained from the bare electronic bands (\ref{eq:bare_fermionic_hamiltonian}): 
\begin{align}\label{eq:barebubble}
\chi^{(0)}_{\rho \rho, \sigma} ({\bm q},i \Omega_n) &=  \frac{1}{\cal A}\sum_{\bm k}\sum_{\tau,\tau^\prime, \lambda,\lambda^\prime}B^{\lambda, \lambda^\prime; \tau, \tau^\prime}_{\sigma}({\bm k}, {\bm q}) \nonumber\\
&\times \frac{n_{\rm F} (E_{{\bm k},\lambda^\prime,\sigma}) - n_{\rm F} (E_{{\bm k}+{\bm q},\lambda,\sigma})}{i \Omega_n - E_{{\bm k}+{\bm q},\lambda,\sigma} + E_{{\bm k},\lambda^\prime,\sigma}}~.
\end{align}
Here, $\cal A$ is the 2D system's area, $B^{\lambda, \lambda^\prime; \tau, \tau^\prime}_{\sigma}({\bm k}, {\bm q}) \equiv \Gamma_{\tau,\lambda^\prime,\sigma}^{\dagger} ({\bm k}) \Gamma_{\tau,\lambda,\sigma} ({\bm k}+{\bm q}) \Gamma^\dagger_{\tau^\prime,\lambda,\sigma}({\bm k}+{\bm q}) \Gamma_{\tau^\prime,\lambda^\prime,\sigma}({\bm k})$ and $n_{\rm F} (x) = 1/ (e^{\beta(x - \mu)} + 1)$ is the Fermi-Dirac distribution function, $T=(k_{\rm B}\beta)^{-1}$ and $\mu$ being the temperature and chemical potential~\cite{Giuliani2005}. In writing Eq.~(\ref{eq:barebubble}) and in all the equations below we have set $\hbar = 1$.  

\begin{figure}[t]
\begin{overpic}[width=0.90\columnwidth]{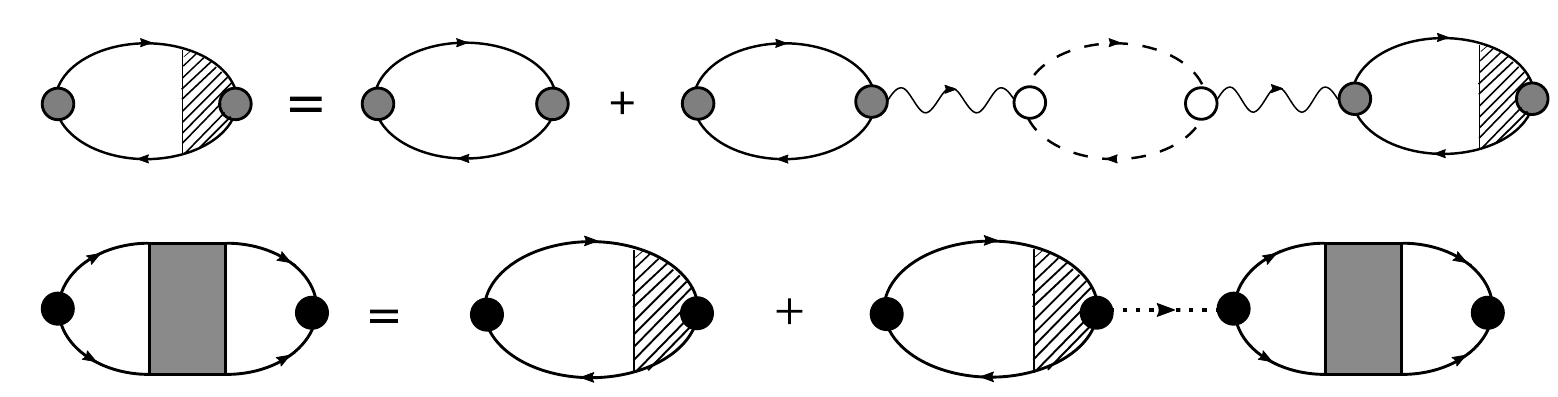} 
\put(-1,55){\rm{(a)}} 
\put(-1,25){\rm{(b)}}
\end{overpic}
\caption{(a) Diagrammatic representation of the Dyson equation for the density-density response function including electron-magnon interactions at the RPA level. The vertices are same as in Fig.~\ref{fig:second_order}. (b) Dyson series expansion of density-density response function including electron-electron interaction at the RPA level. Here, the filled black circles refer to the electron density operators $\hat{\rho} ({\bm q}) = \sum_\sigma \hat{\rho}_\sigma({\bm q})$.
\label{fig:em-ee}}
\end{figure}

We now proceed to calculate the density-density response function including electron-magnon interactions. To this end, we first note such response function can be decomposed into spin channels since the electron-magnon interaction $\hat{\cal H}^\prime_{1, {\rm em}}$ in Eq.~(\ref{Eq:el-mag}) does not mix them. This is not a consequence of symmetry, but just a result of our perturbative Schrieffer-Wolf calculation, which neglects terms ${\cal O}(1/S)$~\cite{Nagaev1974}. We are therefore led to define the spin-resolved density-density response function including electron-magnon interactions, i.e.~$\chi^{\rm (em)}_{\rho \rho, \sigma}({\bm q}, i\Omega_n)$. We carry out a perturbative expansion in the parameter $\tilde{\cal I}$. The lowest order terms that contribute to the expansion are of second order in the $\tilde{\cal I}$ and are shown in Fig.~\ref{fig:second_order}. 
For example, Fig.~\ref{fig:second_order}(a) is a vertex correction while (b) and (c) are self-energy diagrams. These are {\it proper} diagrams, which cannot be split into two disconnected pieces by cutting out a single magnon line. They contribute to the {\it proper} density-density response function. Clearly, the diagram in Fig.~\ref{fig:second_order}(d) is an {\it improper} diagram and does not contribute to the proper density-density response.

In the following, we carry out the {\it first} RPA, in which the proper density-density response is replaced by the non-interacting one $\chi^{(0)}_{\rho \rho, \sigma}$. The resulting spin-resolved density-density response function including electron-magnon interactions ($\chi^{\rm (em)}_{\rho \rho, \sigma}$) at the RPA level is obtained by summing the infinite set of diagrams shown in Fig.~\ref{fig:em-ee}(a). We find:
\begin{align}\label{Eq_double_RPA} 
 \chi^{\rm (em)}_{{\rho \rho},\sigma} ({\bm q},i\Omega_n) &- \sigma \gamma_\sigma ({\bm q},i\Omega_n) \sum_{\sigma^\prime} \sigma^\prime  \chi^{\rm (em)}_{{\rho \rho},\sigma^\prime} ({\bm q},i\Omega_n)\nonumber\\
 & = \chi^{(0)}_{{\rho \rho},\sigma} ({\bm q},i\Omega_n)~,
\end{align}
where $\gamma_\sigma ({\bm q},i\Omega_n)$ is given by
\begin{align}
\gamma_\sigma (&{\bm q},i\Omega_n) =\frac{1}{{\cal A}^2} \sum_{{\bm k},{\bm q^\prime}} \sum_{\tau, \tau^\prime} \sum_{\lambda,\lambda^\prime} \sum_{\zeta, \zeta^\prime} \tilde{\cal I}^{\lambda, \lambda^\prime}_{{\bm k},{\bm q^\prime},-{\bm q}} \tilde{\cal I}^{\lambda^\prime, \lambda}_{{\bm k}+{\bm q},{\bm q^\prime}-{\bm q},{\bm q}} \nonumber\\
&\times A^{\zeta, \zeta^\prime; \lambda^\prime, \lambda}_{\tau, \sigma}({\bm k}, {\bm q}^\prime, -{\bm q})A^{\zeta^\prime, \zeta; \lambda, \lambda^\prime}_{\tau^\prime, \sigma}({\bm k} + {\bm q}, {\bm q}^\prime - {\bm q}, {\bm q})
\nonumber\\
&\times 
\frac{n_{\rm F} (E_{{\bm k},\lambda^\prime,\sigma})-n_{\rm F} (E_{{\bm k}+{\bm q},\lambda,\sigma})}{i\Omega_n - E_{{\bm k}+{\bm q},\lambda,\sigma} + E_{{\bm k},\lambda^\prime,\sigma}} \nonumber\\
&\times
\frac{n_{\rm B} (\Omega^{\rm (m)}_{{\bm q^\prime}-{\bm q},\zeta^\prime}) - n_{\rm B} (\Omega^{\rm (m)}_{{\bm q^\prime},\zeta})}{i\Omega_n - \Omega^{\rm (m)}_{{\bm q^\prime},\zeta} + \Omega^{\rm (m)}_{{\bm q^\prime}-{\bm q},\zeta^\prime}}~.
\label{Eq_gamma}
\end{align}
Here, $n_{\rm B} (x) = 1/ (e^{\beta x} - 1)$ is the Bose-Einstein distribution function.

Solving Eq.~(\ref{Eq_double_RPA}) and summing over both spin components we finally find:
\begin{widetext}
\begin{equation}
\chi^{\rm (em)}_{{\rho \rho}} ({\bm q},i\Omega_n) = \frac{\chi^{(0)}_{{\rho \rho}} ({\bm q},i\Omega_n) - 2 \big[\gamma_\downarrow ({\bm q},i\Omega_n) \chi^{(0)}_{{\rho \rho},\uparrow} ({\bm q},i\Omega_n) + \gamma_\uparrow ({\bm q},i\Omega_n) \chi^{(0)}_{{\rho \rho},\downarrow} ({\bm q},i\Omega_n)\big]}{1 - [\gamma_\uparrow ({\bm q},i\Omega_n) + \gamma_\downarrow ({\bm q},i\Omega_n)]}~
\end{equation}
\end{widetext}
where $\chi^{\rm (em)}_{{\rho \rho}} ({\bm q},i\Omega_n) \equiv \sum_\sigma \chi^{\rm (em)}_{\rho \rho, \sigma} ({\bm q},i\Omega_n)$ and $\chi^{(0)}_{\rho \rho} ({\bm q},i \Omega_n) \equiv \sum_\sigma \chi^{(0)}_{{\rho \rho}, \sigma} ({\bm q},i \Omega_n)$.

So far, the two itinerant electron subsystems, one of electrons with spin $\uparrow$ and one of electrons with spin $\downarrow$, are decoupled. We now include electron-electron interactions into the picture, which couple the two fluids, arriving at the approximate density-density response function $\chi_{\rho \rho} ({\bm q},i\Omega_n)$ that is needed in Eqs.~(\ref{Eq:loss})-(\ref{eq:relation_epsilon_chi}). To this end, we perform a {\it second} RPA, where we now sum the infinite sum of diagrams shown in Fig.~\ref{fig:em-ee}(b), the dotted line denoting the Coulomb interaction $v_{\bm q}$. Carrying out the summation, we find $\chi_{\rho \rho} ({\bm q},i\Omega_n) = \chi^{({\rm em})}_{\rho \rho} ({\bm q},i\Omega_n)/ [1-v_{\bm q} \chi^{({\rm em})}_{\rho \rho} ({\bm q},i\Omega_n)]$. The dielectric function in the double RPA formalism is then given as
\begin{equation}\label{eq:final_result_electron_electron_electron_magnon}
\varepsilon ({\bm q},i\Omega_n) =1 - v_{\bm q} \chi^{({\rm em})}_{\rho \rho} ({\bm q},i\Omega_n)~.
\end{equation}
It should be noted that the order in which the two RPAs are carried out does not change the final result (\ref{eq:final_result_electron_electron_electron_magnon}).

\subsection{Numerical results}

An illustrative numerical result for the loss function \eqref{Eq:loss}, resulting from the double RPA dynamical dielectric function in Eq.~(\ref{eq:final_result_electron_electron_electron_magnon}), is reported in Fig.~\ref{fig:loss}, for wave vectors ${\bm q}$ near the $\Gamma$ point.  In the absence of  electron-magnon interactions, we have an ordinary long-lived plasmon (white line in Fig.~\ref{fig:loss}), which in the long-wavelength limit is well described by the usual 2D plasmon formula  $\Omega_{\rm pl}(q\to 0) = [2 \pi n e^2 q/(\kappa m_{\rm b})]^{1/2}$, where $m_{\rm b}$ is the carrier band mass and $n$ is the carrier concentration~\cite{Giuliani2005}. When  electron-magnon interactions are turned on, the loss function displays an important feature akin to an avoided crossing at the energy at which the bare plasmon $\Omega_{\rm pl}$ meets the upper magnon branch. In condensed matter systems we are used to avoided crossings between e.g.~bare photons and matter modes such as phonons and excitons, yielding polaritons~\cite{Basov_Nanophotonics_2021}. This type of hybridization stems from a coupling of the type $(\hat{a}+\hat{a}^\dagger)(\hat{b} + \hat{b}^\dagger )$, where $\hat{a}^\dagger$ ($\hat{a}$) are photonic creation (annihilation) operators while $\hat{b}^\dagger$ ($\hat{b}$) creates (annihilates) a bosonic mode of the matter degrees of freedom. Here, instead, coupling between magnons and plasmons arises from the microscopic exchange interaction (\ref{electron-spin}) between localized spins and itinerant electrons and from the long-range Coulomb interactions in Eq.~(\ref{eq:Coulomb_general}) between itinerant electrons.

For the parameters used in Fig.~\ref{fig:loss}, which are relevant to ${\rm Cr}_2{\rm Ge}_2{\rm Te}_6$, such feature occurs at frequencies on the order of $20~{\rm THz}$ and is therefore fully measurable. 

\section{Summary and conclusion}
\label{sect:conclusions}
In summary, we have shown that plasmon-magnon interactions clearly manifest in the spectra of collective charge excitations of 2D doped magnetic materials, yielding a splitting of the plasmon dispersion into two branches---Fig.~\ref{fig:loss}.  Such splitting can therefore be probed by cryogenic SNOM~\cite{Yang2013,Ni2018,Chen2019,Sun2020}. This is important since standard neutron scattering techniques cannot be used to investigate the properties of spin waves in 2D materials. 

Once again, we hasten to emphasize that, in stark contrast to common wisdom~\cite{Barnas1979,Efimkin2021,Dyrdal2022}, plasmon-magnon interactions do not require spin-orbit coupling to exist. They can arise solely from the exchange interaction between the itinerant carriers and the localized magnetic moments hosted by the lattice. To this end, in the future it will be interesting to set up a macroscopic theoretical approach to such coupled plasmon-magnon dynamics by combining an Euler equation for the charge density~\cite{Giuliani2005} with a suitable Landau-Lifshitz-Gilbert equation for the magnetization density.

Even though this Article was focussed, for the sake of concreteness,  on monolayer ${\rm Cr}_2{\rm Ge}_2{\rm Te}_6$, our formalism is readily applicable to other 2D magnetic materials~\cite{Gibertini2019}.

\section*{Acknowledgements}
We thank Amir Yacoby for useful and inspiring discussions.
S.G. acknowledges hospitality from TU Dresden during the first wave of the Covid-19 pandemic. 

This work was supported by the European Union's Horizon 2020 research and innovation programme under grant agreement No.~881603 - GrapheneCore3 and under the Marie Sklodowska-Curie grant agreement No.~873028 - Hydrotronics, by the University of Pisa under the ``PRA - Progetti di Ricerca di Ateneo'' (Institutional Research Grants) -  Project No.~PRA\_2020-2021\_92~``Quantum Computing, Technologies and Applications'', and by the MUR - Italian Minister of University and Research under the ``Research projects of relevant national interest  - PRIN 2020''  - Project No.~2020JLZ52N, title ``Light-matter interactions and the collective behavior of quantum 2D materials (q-LIMA)''.  The work of M.I.K. was supported by the ERC Synergy Grant, Project No.~854843 ``FASTCORR''.


\appendix

%
\section{Equation-of-motion approach for the calculation of the dielectric function} 
\label{SM_eq_of_motion}

To validate the diagrammatic double RPA method of calculating the dielectric function presented in the main text, we here present a calculation of $\varepsilon({\bm q}, \Omega)$ based on the equation of motion approach. This method is also known as the method of ``self-consistent fields''~\cite{Mahan2000,Bruus2004}. The essential idea of this approach is to introduce an external charge density $\rho_{\rm ext}({\bm r},t) $ and its  Fourier transform $\rho_{\rm ext}({\bm q},\Omega)$. Such excess charge  is screened by mobile electrons and the total (a.k.a. ``screened potential'') potential $V$ is the sum of the external $V_{\rm ext}$ and induced $V_{\rm ind}$ potentials:
\begin{equation}
V({\bm q},\Omega) = V_{\rm ext}({\bm q},\Omega) + V_{\rm ind}({\bm q},\Omega)~.
\label{app_eq_V}
\end{equation}
Once $V ({\bm q},\Omega)$ is found out, the dielectric function $\varepsilon({\bm q}, \Omega)$ can be calculated by $\varepsilon ({\bm q},\Omega) = V_{\rm ext} ({\bm q},\Omega) / V ({\bm q},\Omega)$.

The effective Hamiltonian of our system reads
\begin{equation}
\hat{\cal H}^\prime = \hat{\cal H}^\prime_{e} + \hat{\cal H}^\prime_{\rm s} + \hat{\cal H}^\prime_{1,\rm em} + \frac{1}{\cal A}\sum_{\bm q} V({\bm q},t) \hat{\rho}({\bm q})~, 
\label{App_Eq_hamil}
\end{equation}
where the electronic density operator is expressed, using Eq.~\eqref{Eq:unitary_electron}, as
\begin{align}\label{eq:Eq_density_op}
\hat{\rho} ({\bm q}) = &\sum_{{\bm k},\tau,\sigma}  \hat{c}^\dagger_{{\bm k}+{\bm q},\tau,\sigma} \hat{c}_{{\bm k},\tau,\sigma}  \nonumber \\ = &  \sum_{{\bm k},\tau,\sigma,\lambda,\lambda^\prime}  \Gamma^\dagger_{\tau,\lambda,\sigma}({\bm k} + {\bm q})\Gamma_{\tau,\lambda^\prime,\sigma}({\bm k}) \nonumber \\ & \times \hat{f}^{\dagger}_{{\bm k} + {\bm q},{\lambda},\sigma} \hat{f}_{{\bm k},{\lambda^\prime}\sigma}~,
\end{align}
and the 2D system's area ${\cal A}$ has been introduced in the main text.

It is important to note that there is no explicit electron-electron Coulomb repulsion term in the Hamiltonian. It is included indirectly in the Hamiltonian through the total potential $V$. In fact, the induced part $V_{\rm ind}$ arises from the Coulomb interaction~\cite{Giuliani2005}. 

We now write down the equation of motion for the density operator, i.e.
\begin{equation}
\frac{d}{dt} \hat{\rho} ({\bm q},t) = i [\hat{\cal H}^\prime,\hat{\rho} ({\bm q},t)]~.
\label{Eq:density}
\end{equation}
The external charge creates a classical oscillating field, and we are trying to find the quantum response of the electron gas to this classical oscillation. Furthermore, since the external field is assumed to be oscillating at a frequency $\Omega$, the density response of the electron system will also depend on $\Omega$. Therefore, we have $\hat{\rho} ({\bm q},t) \sim \hat{\rho} ({\bm q}) e^{i \Omega t}$. 

Rather than Eq.~\eqref{Eq:density}, it is more convenient to consider the equation of motion for the operator $ \hat{f}^\dagger_{{\bm k}+{\bm q},\lambda,\sigma} \hat{f}_{{\bm k},\lambda^\prime,\sigma}$:
\begin{equation}\label{Eq_motion}
\frac{d}{dt}  \hat{f}^\dagger_{{\bm k}+{\bm q},\lambda,\sigma} \hat{f}_{{\bm k},\lambda^\prime,\sigma} = i [\hat{\cal H}^\prime,\hat{f}^\dagger_{{\bm k}+{\bm q},\lambda,\sigma} \hat{f}_{{\bm k},\lambda^\prime,\sigma}]~.
\end{equation}
Let us first investigate the case in which $\hat{\cal H}^\prime_{1,\rm em} = 0$.
In what follows, we list all the commutators that are needed to evaluate the right-hand side of Eq.~(\ref{Eq_motion}):
\begin{widetext}
\begin{align}
[\hat{\cal H}^\prime_{\rm e},\hat{f}^\dagger_{{\bm k}+{\bm q},\lambda,\sigma} \hat{f}_{{\bm k},\lambda^\prime,\sigma}] =& \left( E_{{\bm k}+{\bm q},\lambda,\sigma} - E_{{\bm k},\lambda^\prime,\sigma}\right) \hat{f}^\dagger_{{\bm k}+{\bm q},\lambda,\sigma}  \hat{f}_{{\bm k},\lambda^\prime,\sigma}~,\\
[\hat{\cal H}^\prime_{\rm s},\hat{f}^\dagger_{{\bm k}+{\bm q},\lambda,\sigma} \hat{f}_{{\bm k},\lambda^\prime,\sigma}] =&~0~, \\
\sum_{\bm q^\prime} V ({\bm q^\prime},t) [\hat{\rho}({\bm q^\prime}),\hat{f}^\dagger_{{\bm k}+{\bm q},\lambda,\sigma} \hat{f}_{{\bm k},\lambda^\prime,\sigma}] =& \sum_{\bm q^\prime,\tau} V ({\bm q^\prime},t) \left[ \Gamma_{\tau,\lambda^\prime,\sigma}^{\dagger} ({\bm k}+{\bm q}+{\bm q^\prime}) \Gamma_{\tau,\lambda,\sigma} ({\bm k}+{\bm q}) \hat{f}^\dagger_{{\bm k}+{\bm q}+{\bm q^\prime},\lambda^\prime,\sigma} \hat{f}_{{\bm k},\lambda^\prime,\sigma} \right.\nonumber \\ 
& - \left.\Gamma_{\tau,\lambda^\prime,\sigma}^{\dagger} ({\bm k}) \Gamma_{\tau,\lambda,\sigma} ({\bm k}-{\bm q^\prime}) \hat{f}^\dagger_{{\bm k}+{\bm q},\lambda,\sigma} \hat{f}_{{\bm k}-{\bm q^\prime},\lambda,\sigma}  \right]~.
\end{align}
\end{widetext}

The right hand side of the last equation is approximated by retaining only the term with ${\bm q^\prime}= - {\bm q}$ in the summation. Terms with other values of ${\bm q^\prime}$ are assumed to average out to zero. In this formalism, it is very well known that including only the contribution with. ${\bm q^\prime}= - {\bm q}$ is equivalent to the RPA~\cite{Mahan2000,Bruus2004}. We also assume $V (- {\bm q},t) = V({\bm q},t)$. 

Under these assumptions, the equation of motion reads
\begin{align}
\left( \Omega \right. & \left.- E_{{\bm k}+{\bm q},\lambda,\sigma}  +  E_{{\bm k},\lambda^\prime,\sigma}\right) \hat{f}^\dagger_{{\bm k}+{\bm q},\lambda,\sigma} \hat{f}_{{\bm k},\lambda^\prime,\sigma} = \nonumber \\  & =  V({\bm q},t) \frac{1}{\cal A} \sum_\tau \Gamma_{\tau,\lambda^\prime,\sigma}^{\dagger} ({\bm k}) \Gamma_{\tau,\lambda,\sigma}({\bm k}+{\bm q}) \nonumber \\ &\times  \left(\hat{f}^\dagger_{{\bm k},\lambda^\prime,\sigma} \hat{f}_{{\bm k},\lambda^\prime,\sigma} - \hat{f}^\dagger_{{\bm k}+{\bm q},\lambda,\sigma} \hat{f}_{{\bm k}+{\bm q},\lambda,\sigma} \right)~. 
\end{align}
Therefore, summing over ${\bm k},\sigma, \tau, \tau^\prime, \lambda, \lambda^\prime$, the density operator---see Eq.~(\ref{eq:Eq_density_op})---can be written as:
\begin{align}
\hat{\rho} ({\bm q}) =& \frac{1}{\cal A}V({\bm q},t) \sum_{\sigma}  \sum_{\bm k}\sum_{\tau,\tau^\prime, \lambda,\lambda^\prime} B^{\lambda, \lambda^\prime; \tau, \tau^\prime}_{\sigma}({\bm k}, {\bm q})\nonumber \\ & \times \frac{\hat{f}^\dagger_{{\bm k},\lambda^\prime,\sigma} \hat{f}_{{\bm k},\lambda^\prime,\sigma} - \hat{f}^\dagger_{{\bm k}+{\bm q},\lambda,\sigma} \hat{f}_{{\bm k}+{\bm q},\lambda,\sigma}}{\Omega - E_{{\bm k}+{\bm q},\lambda,\sigma} + E_{{\bm k},\lambda^\prime,\sigma} + i0^+}~.
\end{align}
Here, $B^{\lambda, \lambda^\prime; \tau, \tau^\prime}_{\sigma}({\bm k}, {\bm q}) \equiv \Gamma_{\tau,\lambda^\prime,\sigma}^{\dagger} ({\bm k}) \Gamma_{\tau,\lambda,\sigma} ({\bm k}+{\bm q}) \Gamma^\dagger_{\tau^\prime,\lambda,\sigma}({\bm k}+{\bm q}) \Gamma_{\tau^\prime,\lambda^\prime,\sigma}({\bm k})$.
We now take the average of this equation, so that $\langle \hat{\rho}({\bm q}, t) \rangle$ becomes $ \rho ({\bm q},\Omega) e^{i \Omega t}$, while  $\langle \hat{f}^\dagger_{{\bm k}+{\bm q},\lambda,\sigma} \hat{f}_{{\bm k}+{\bm q},\lambda,\sigma} \rangle$ and $\langle \hat{f}^\dagger_{{\bm k},\lambda^\prime,\sigma} \hat{f}_{{\bm k},\lambda^\prime,\sigma}\rangle$ are replaced by the Fermi-Dirac distribution functions $n_{\rm F} (E_{{\bm k}+{\bm q},\lambda,\sigma})$ and $n_{\rm F} (E_{{\bm k},\lambda^\prime,\sigma})$, respectively. 
Finally, we take $V({\bm q}, t) =V ({\bm q},\Omega) e^{i \Omega t}$. Carrying out these steps, we find:
\begin{widetext}
\begin{equation}
\rho ({\bm q},\Omega) = V({\bm q},\Omega)\underbrace{\frac{1}{\cal A}\sum_{\sigma} \sum_{\bm k}\sum_{\tau,\tau^\prime, \lambda,\lambda^\prime} B^{\lambda, \lambda^\prime; \tau, \tau^\prime}_{\sigma}({\bm k}, {\bm q}) \frac{n_{\rm F} (E_{{\bm k},\lambda^\prime,\sigma}) - n_{\rm F} (E_{{\bm k}+{\bm q},\lambda,\sigma})}{\Omega - E_{{\bm k}+{\bm q},\lambda,\sigma} + E_{{\bm k},\lambda^\prime,\sigma}+i 0^+}}_{\chi^{(0)}_{\rho\rho}({\bm q}, \Omega)}~.
\label{eq:rhoeqmotion}
\end{equation}
\end{widetext}

The quantity relating $\rho({\bm q}, \Omega)$ to $V({\bm q},\Omega)$ is easily seen to coincide with the polarization operator or density-density response function~\cite{Mahan2000,Bruus2004,Giuliani2005} $\chi^{(0)}_{\rho\rho}({\bm q}, \Omega) = \chi^{(0)}_{\rho\rho, \uparrow}({\bm q},\Omega)+ \chi^{(0)}_{\rho\rho, \downarrow}({\bm q}, \Omega)$, where the spin-resolved quantities $\chi^{(0)}_{\rho\rho, \sigma}({\bm q}, \Omega)$ have been introduced in Eq.~(\ref{eq:barebubble}) of the main text.

We now investigate the effect of the electron-magnon interaction by including the commutator $[\hat{\cal H}^\prime_{1,\rm em},\hat{f}^\dagger_{{\bm k}+{\bm q},\lambda,\sigma} \hat{f}_{{\bm k},\lambda^\prime,\sigma}]$ in the equation of motion (\ref{Eq_motion}). The commutator is given by:
\begin{widetext}
\begin{align}
[\hat{\cal H}^\prime_{1,\rm em},\hat{f}^\dagger_{{\bm k}+{\bm q},\lambda,\sigma} \hat{f}_{{\bm k},\lambda^\prime,\sigma}] =& \sum_{{\bm k^\prime},{\bm q^\prime},{\bm p^\prime}} \sum_{\tau} \sum_{\lambda_1 \lambda_2} \sum_{\zeta_1, \zeta_2} \sum_{\sigma^\prime} \sigma^\prime  \tilde{\cal I}^{\lambda_1, \lambda_2}_{{\bm k^\prime},{\bm q^\prime},{\bm p^\prime}}~  A^{\zeta_1, \zeta_2; \lambda_1, \lambda_2}_{\tau, \sigma^\prime}({\bm k^\prime}, {\bm q^\prime}, {\bm p^\prime})
\nonumber \\
&\times\hat{\mu}_{{\bm q^\prime},\zeta_1}^\dagger \hat{\mu}_{{\bm q^\prime}+{\bm p^\prime},\zeta_2} [\hat{f}^{\dagger}_{{\bm k^\prime},{\lambda_1},\sigma^\prime}\hat{f}_{{\bm k^\prime}-{\bm p^\prime},{\lambda_2},\sigma^\prime},\hat{f}^\dagger_{{\bm k}+{\bm q},\lambda,\sigma} \hat{f}_{{\bm k},\lambda^\prime,\sigma}] \nonumber \\
=& \sum_{{\bm q^\prime}} \sum_{\tau} \sum_{\zeta_1, \zeta_2} \sigma \tilde{\cal I}^{\lambda^\prime, \lambda}_{{\bm k},{\bm q^\prime},-{\bm q}}~ A^{\zeta_1, \zeta_2; \lambda^\prime, \lambda}_{\tau, \sigma}({\bm k}, {\bm q^\prime}, -{\bm q}) 
\nonumber \\
&\times \hat{\mu}_{{\bm q^\prime},\zeta_1}^\dagger \hat{\mu}_{{\bm q^\prime}-{\bm q},\zeta_2} \left( \hat{f}^{\dagger}_{{\bm k},{\lambda^\prime},\sigma}\hat{f}_{{\bm k},{\lambda^\prime},\sigma} - \hat{f}^\dagger_{{\bm k}+{\bm q},\lambda,\sigma} \hat{f}_{{\bm k}+{\bm q},\lambda,\sigma}\right)~,
\label{eq:eq_motion_em}
\end{align}
\end{widetext}
where $A^{\zeta, \zeta^\prime; \lambda, \lambda^\prime}_{\tau, \sigma}({\bm k}, {\bm q}, {\bm p})$ has been introduced in Eq.~(\ref{Eq:el-mag_A}) in the main text. In the last step, similarly to the previous case with just the Coulomb interaction, we used the RPA to carry out the sum over ${\bm p^\prime}$ by neglecting all the terms other than the one with ${\bm p^\prime} = -{\bm q}$.

Looking at Eq.~(\ref{eq:eq_motion_em}) it is evident that we need to write down also the equation of motion for the magnon number operators $\hat{\mu}_{{\bm q^\prime},\zeta_1}^\dagger \hat{\mu}_{{\bm q^\prime}-{\bm q},\zeta_2}$. This will involve again the equation of motion for the electron number operators, thereby allowing us to solve them self-consistently. Analyzing different terms of the Hamiltonian, we get:
\begin{widetext}
\begin{align}
[\hat{\cal H}^\prime_{\rm e},\hat{\mu}_{{\bm q^\prime},\zeta_1}^\dagger \hat{\mu}_{{\bm q^\prime}-{\bm q},\zeta_2}] =& 0~,\\
 \sum_{\bm q^\prime} V ({\bm q^\prime},t) [\hat{\rho}({\bm q^\prime}),\hat{\mu}_{{\bm q^\prime},\zeta_1}^\dagger \hat{\mu}_{{\bm q^\prime}-{\bm q},\zeta_2}] =& 0~,\\
[\hat{\cal H}^\prime_{\rm s},\hat{\mu}_{{\bm q^\prime},\zeta_1}^\dagger \hat{\mu}_{{\bm q^\prime}-{\bm q},\zeta_2}] =&\left[\Omega^{\rm (m)}_{{\bm q^\prime},\zeta_1} - \Omega^{\rm (m)}_{{\bm q^\prime}-{\bm q},\zeta_2}\right] \hat{\mu}_{{\bm q^\prime},\zeta_1}^\dagger \hat{\mu}_{{\bm q^\prime}-{\bm q},\zeta_2} \\
[\hat{\cal H}^\prime_{1,\rm em},\hat{\mu}_{{\bm q^\prime},\zeta_1}^\dagger \hat{\mu}_{{\bm q^\prime}-{\bm q},\zeta_2}] =& \sum_{{\bm k^{\prime\prime}},{\bm q^{\prime\prime}},{\bm p^{\prime\prime}}} \sum_{\tau^\prime} \sum_{\lambda_1 \lambda_2} \sum_{\zeta_3, \zeta_4} \sum_{\sigma^{\prime\prime}} \sigma^{\prime\prime}  \tilde{\cal I}^{\lambda_1, \lambda_2}_{{\bm k^{\prime\prime}},{\bm q^{\prime\prime}},{\bm p^{\prime\prime}}}~A^{\zeta_3, \zeta_4; \lambda_1, \lambda_2}_{\tau^\prime, \sigma^{\prime\prime}}({\bm k^{\prime\prime}}, {\bm q^{\prime\prime}}, {\bm p^{\prime\prime}}) 
\nonumber \\
&\times \hat{f}^{\dagger}_{{\bm k^{\prime\prime}},{\lambda_1},\sigma^{\prime\prime}}\hat{f}_{{\bm k^{\prime\prime}}-{\bm p^{\prime\prime}},{\lambda_2},\sigma^{\prime\prime}} [\hat{\mu}_{{\bm q^{\prime\prime}},\zeta_3}^\dagger \hat{\mu}_{{\bm q^{\prime\prime}}+{\bm p^{\prime\prime}},\zeta_4},\hat{\mu}_{{\bm q^{\prime}},\zeta_1}^\dagger \hat{\mu}_{{\bm q^\prime}-{\bm q},\zeta_2}]~, \nonumber \\
=& \sum_{\tau^\prime} \sum_{\sigma^{\prime\prime}} \sum_{\lambda_1, \lambda_2} \sigma^{\prime\prime} \tilde{\cal I}^{\lambda_1, \lambda_2}_{{\bm k}+{\bm q},{\bm q^\prime}-{\bm q},{\bm q}}~A^{\zeta_2, \zeta_1; \lambda_1, \lambda_2}_{\tau^\prime, \sigma^{\prime\prime}}({\bm k}+{\bm q}, {\bm q^\prime}-{\bm q}, {\bm q}) \nonumber \\ 
&\times \hat{f}^{\dagger}_{{\bm k}+{\bm q},\lambda_1,\sigma^{\prime\prime}}\hat{f}_{{\bm k},\lambda_2,\sigma^{\prime\prime}} \left( \hat{\mu}_{{\bm q^\prime}-{\bm q},\zeta_2}^\dagger \hat{\mu}_{{\bm q^\prime}-{\bm q},\zeta_2} - \hat{\mu}_{{\bm q^\prime},\zeta_1}^\dagger \hat{\mu}_{{\bm q^\prime},\zeta_1} \right)~.
\end{align}
\end{widetext}

Here, we have again used the RPA as in Eq.~(\ref{eq:eq_motion_em}),  neglecting all the terms of the momentum sum other than the term with ${\bm p^{\prime\prime}} = -{\bm q}$. As described previously, we take the average of the operators assuming a time dependence of the form $e^{i \Omega t}$ and replace the magnon number operators with Bose-Einstein distribution functions $n_{\rm B}$. We find
\begin{align}
\hat{\mu}_{{\bm q^\prime},\zeta_1}^\dagger \hat{\mu}_{{\bm q^\prime}-{\bm q},\zeta_2} &= \sum_{\tau^\prime} \sum_{\sigma^{\prime\prime}} \sum_{\lambda_1, \lambda_2} \sigma^{\prime\prime}  \tilde{\cal I}^{\lambda_1, \lambda_2}_{{\bm k}+{\bm q},{\bm q^\prime}-{\bm q},{\bm q}}\nonumber \\ &\times A^{\zeta_2, \zeta_1; \lambda_1, \lambda_2}_{\tau^\prime, \sigma^{\prime\prime}}({\bm k}+{\bm q}, {\bm q^\prime}-{\bm q}, {\bm q})\nonumber \\ &\times \hat{f}^{\dagger}_{{\bm k}+{\bm q},{\lambda_1},\sigma^{\prime\prime}}\hat{f}_{{\bm k},{\lambda_2},\sigma^{\prime\prime}} \nonumber \\
&\times \frac{n_{\rm B} (\Omega^{\rm (m)}_{{\bm q^\prime}-{\bm q},\zeta_2}) -  n_{\rm B} (\Omega^{\rm (m)}_{{\bm q^\prime},\zeta_1})}{\Omega - \Omega^{\rm (m)}_{{\bm q^\prime},\zeta_1} + \Omega^{\rm (m)}_{{\bm q^\prime}-{\bm q},\zeta_2} + i 0^+}. 
\end{align}
Using this result in Eq.~(\ref{Eq_motion}) we finally find
\begin{widetext}
\begin{align}
&(\Omega - E_{{\bm k}+{\bm q},\lambda,\sigma} + E_{{\bm k},\lambda^\prime,\sigma}) \hat{f}^\dagger_{{\bm k}+{\bm q},\lambda,\sigma} \hat{f}_{{\bm k},\lambda^\prime,\sigma} - \sum_{{\bm q^\prime}} \sum_{\tau} \sum_{\zeta_1, \zeta_2} \sigma~ \tilde{\cal I}^{\lambda^\prime, \lambda}_{{\bm k},{\bm q^\prime},-{\bm q}} A^{\zeta_1, \zeta_2; \lambda^\prime, \lambda}_{\tau^\prime, \sigma}({\bm k}, {\bm q^\prime}, -{\bm q}) 
\nonumber \\
&\times \left( \hat{f}^{\dagger}_{{\bm k},{\lambda^\prime},\sigma}\hat{f}_{{\bm k},{\lambda^\prime},\sigma} - \hat{f}^\dagger_{{\bm k}+{\bm q},\lambda,\sigma} \hat{f}_{{\bm k}+{\bm q},\lambda,\sigma}\right) 
\times \sum_{\tau^\prime} \sum_{\sigma^{\prime}} \sum_{\lambda_1, \lambda_2} \sigma^{\prime}  \tilde{\cal I}^{\lambda_1, \lambda_2}_{{\bm k}+{\bm q},{\bm q^\prime}-{\bm q},{\bm q}}~A^{\zeta_2, \zeta_1; \lambda_1, \lambda_2}_{\tau^\prime, \sigma^{\prime}}({\bm k}+{\bm q}, {\bm q^\prime}-{\bm q}, {\bm q})  \nonumber \\
&\times \hat{f}^{\dagger}_{{\bm k}+{\bm q},{\lambda_1},\sigma^{\prime\prime}}\hat{f}_{{\bm k},{\lambda_2},\sigma^{\prime\prime}} \frac{n_{\rm B} (\Omega^{\rm (m)}_{{\bm q^\prime}-{\bm q},\zeta_2}) -  n_{\rm B} (\Omega^{\rm (m)}_{{\bm q^\prime},\zeta_1})}{\Omega - \Omega^{\rm (m)}_{{\bm q^\prime},\zeta_1} + \Omega^{\rm (m)}_{{\bm q^\prime}-{\bm q},\zeta_2}+ i 0^+}= \nonumber \\
&= V({\bm q},t) \frac{1}{\cal A} \sum_\tau \Gamma_{\tau,\lambda^\prime,\sigma}^{\dagger} ({\bm k}) \Gamma_{\tau,\lambda,\sigma} ({\bm k}+{\bm q}) \left(\hat{f}^\dagger_{{\bm k},\lambda^\prime,\sigma} \hat{f}_{{\bm k},\lambda^\prime,\sigma} - \hat{f}^\dagger_{{\bm k}+{\bm q},\lambda,\sigma} \hat{f}_{{\bm k}+{\bm q},\lambda,\sigma} \right)~. 
\end{align}
\end{widetext}
So far we studied the equations of motion for $\hat{f}^\dagger_{{\bm k}+{\bm q},\lambda,\sigma} \hat{f}_{{\bm k},\lambda^\prime,\sigma}$ and $\hat{\mu}_{{\bm q^\prime},\zeta_1}^\dagger \hat{\mu}_{{\bm q^\prime}-{\bm q},\zeta_2}$. Now, recalling Eq.~(\ref{eq:Eq_density_op}), we can write the equation of motion for the spin-resolved density $\rho_{\sigma} ({\bm q},\Omega)$ by summing over repeated indices and taking the averages. We obtain: 
\begin{widetext}
\begin{align}
\rho_{\sigma} ({\bm q},\Omega) - \sigma \gamma_{\sigma} ({\bm q},\Omega) \sum_{\sigma^\prime} \sigma^\prime \rho_{\sigma^\prime} ({\bm q},\Omega) = V ({\bm q},\Omega) \frac{1}{\cal A} \sum_{\bm k}\sum_{\tau,\tau^\prime, \lambda,\lambda^\prime}~B^{\lambda, \lambda^\prime; \tau, \tau^\prime}_{\sigma}({\bm k}, {\bm q}) \frac{n_{\rm F} (E_{{\bm k},\lambda^\prime,\sigma}) - n_{\rm F} (E_{{\bm k}+{\bm q},\lambda,\sigma})}{\Omega - E_{{\bm k}+{\bm q},\lambda,\sigma} + E_{{\bm k},\lambda^\prime,\sigma}+ i 0^+}~,
\label{eq:rhosigma_em}
\end{align}
\end{widetext}
where $\gamma_{\sigma} ({\bm q},\Omega)$ is the same as in Eq.~(\ref{Eq_gamma}) of the main text. We now use the definition of the spin-resolved density-density response function $\chi^{\rm (em)}_{{\rho \rho},\sigma} ({\bm q},\Omega)$ including electron-magnon interactions---see the main text, right before Eq.~(\ref{Eq_double_RPA}):
\begin{equation}\label{eq:def_chi_em}
\rho_{\sigma} ({\bm q},\Omega) = \chi^{\rm (em)}_{{\rho \rho},\sigma} ({\bm q},\Omega)V ({\bm q},\Omega)~.
\end{equation}
Replacing Eq.~(\ref{eq:def_chi_em}) inside Eq.~(\ref{eq:rhosigma_em}) and eliminating $V ({\bm q},\Omega)$ we finally get Eq.~(14) in the main text.

In conclusion, the polarization of mobile electrons caused by the external charge density $\rho_{\rm ext}({\bm q},\Omega)$ results in an induced potential $V_{\rm ind}({\bm q},\Omega) = v_{\bm q} {\rho} ({\bm q},\Omega) =v_{\bm q} \chi^{(\rm em)}_{\rho \rho} ({\bm q},\Omega) V ({\bm q},\Omega)$, where $\chi^{(\rm em)}_{\rho \rho} ({\bm q},\Omega)=\chi^{\rm (em)}_{{\rho \rho},\uparrow}({\bm q},\Omega)+\chi^{\rm (em)}_{{\rho \rho},\downarrow}({\bm q},\Omega)$ and $v_{\bm q}=2\pi e^2/(\kappa q)$ is the 2D Fourier transform of the Coulomb potential~\cite{Giuliani2005}, $\kappa$ being a background dielectric constant. Therefore, from Eq.~\eqref{app_eq_V}, we have the dielectric function which includes both the electron-magnon and Coulomb interactions:
\begin{equation}
\varepsilon ({\bm q},\Omega) = 1 - v_{\bm q} \chi^{(\rm em)}_{\rho \rho} ({\bm q},\Omega)~.
\label{app_eq_varepsilon_2}
\end{equation} 
We conclude that, using the equation of motion method at the RPA level,  we have obtained results that are identical to those obtained from the double RPA diagrammatic approach described in the main text.

\section{Further comments on the role of $\gamma$}
\label{SM_temp}
The calculated electron energy-loss function ${\cal L}({\bm q}, \Omega)$---defined as in Eq.~(11) of the main text---of monolayer doped Cr$_2$Ge$_2$Te$_6$ is plotted in Fig.~1 of the main text. It clearly shows an important feature at the energy at which the 2D bare plasmon dispersion $\Omega_{\rm pl}$ meets the upper magnon branch. 

\begin{figure}[t]
\centering
\begin{tabular}{cc}
\begin{overpic}[unit=1mm,width=0.48\columnwidth]{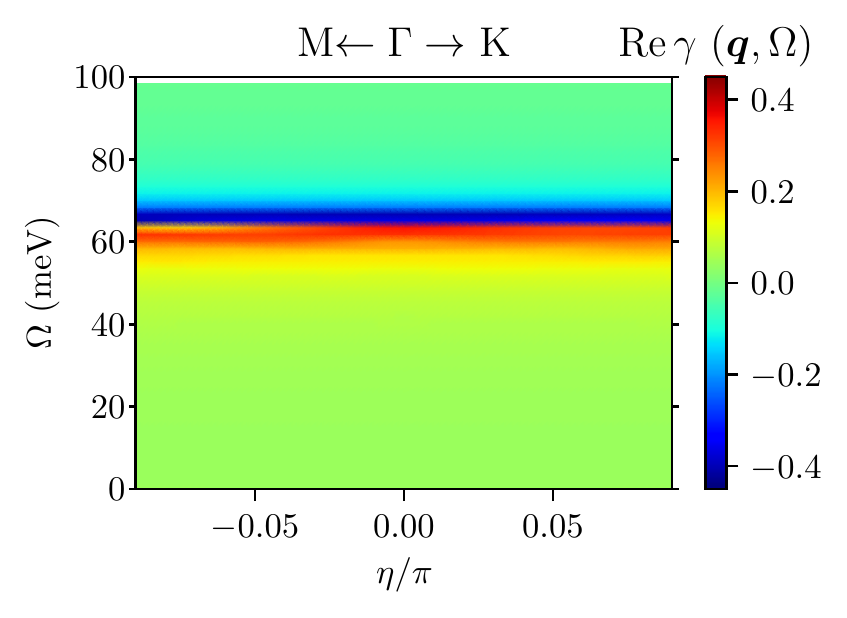}
\put(-1,26){\rm (a)}
\end{overpic} &
\begin{overpic}[unit=1mm,width=0.48\columnwidth]{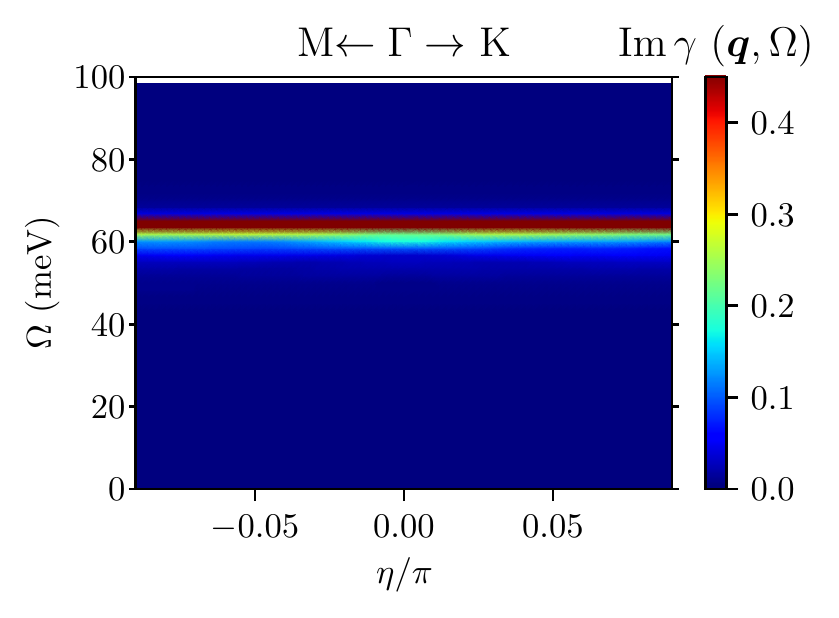}
\put(-1,26){\rm (b)}
\end{overpic} \\
\begin{overpic}[unit=1mm,width=0.48\columnwidth]{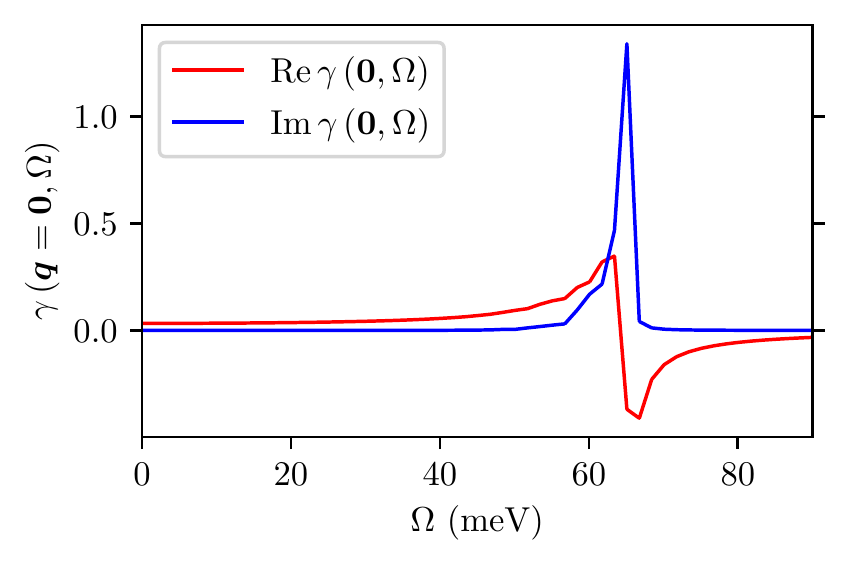}
\put(-1,26){\rm (c)}
\end{overpic} &
\begin{overpic}[unit=1mm,width=0.48\columnwidth]{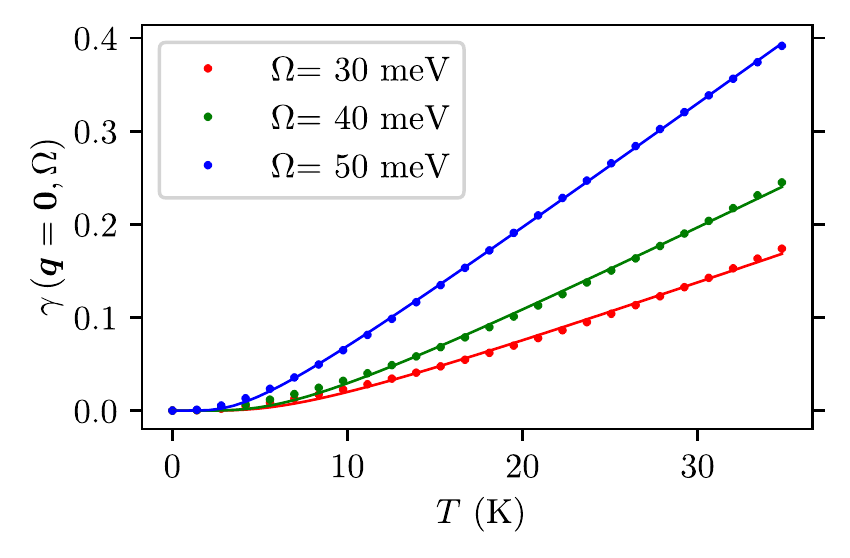}
\put(-1,26){\rm (d)}
\end{overpic} \\
\end{tabular}
\caption{(Color online) Panels (a) and (b) illustrate the real and imaginary parts of $\gamma ({\bm q},\Omega)$, respectively, at $T = 25~{\rm K}$, plotted as functions of ${\bm q}$ and $\Omega$. (c) The real and imaginary part of  $\gamma({\bm q} = {\bm 0},\Omega)$. (d) The temperature dependence of $\gamma ({\bm q}= {\bm 0},\Omega)$ is shown for different values of $\Omega$. Filled circles denote data points from our numerical calculations, while solid lines are fits carried out on the basis of the simple analytical formula (\ref{Eq:gamma}).\label{fig:gamma2}}
\end{figure}

The ${\bm q}$ and of $\Omega$ dependence of the real and imaginary parts of the function $\gamma ({\bm q},\Omega)= \gamma_\uparrow ({\bm q}, \Omega) + \gamma_\downarrow ({\bm q},\Omega)$ at $T=25~{\rm K}$ are reported in panels~(a) and~(b) of Fig.~\ref{fig:gamma2}, respectively. As we can see, $\gamma ({\bm q},\Omega)$ barely depends on ${\bm q}$ and its imaginary part has a divergence at the upper magnon branch ($\approx 58~{\rm meV}$). Fig.~\ref{fig:gamma2}(c) shows in more detail the behavior of $\gamma ({\bm q},\Omega)$ as a function of $\Omega$, for the case of  ${\bm q}={\bm 0}$. As a consequence of such trend $\chi^{(\rm em)}_{{\rho \rho}} ({\bm q},\Omega) \sim \frac{1}{1 - \gamma ({\bm q}, \Omega)}$ vanishes at frequencies just below the upper magnon branch frequency, thereby yielding $\varepsilon ({\bm q},\Omega) \rightarrow 1$ and the quenching of the plasmon oscillation.

\begin{figure}[t]
\centering
\begin{tabular}{cc}
\begin{overpic}[unit=1mm,width=0.48\columnwidth]{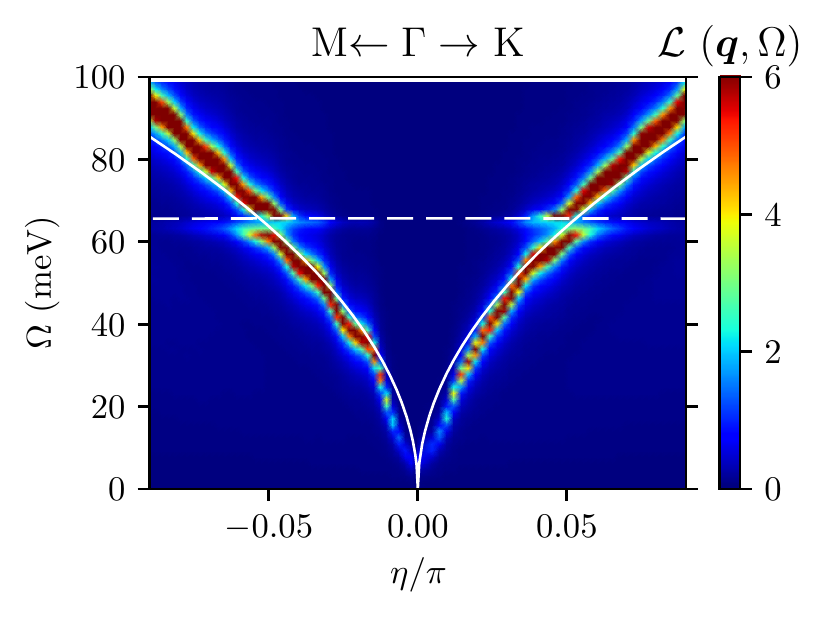}
\put(-1,26){\rm (a)}
\end{overpic} &
\begin{overpic}[unit=1mm,width=0.48\columnwidth]{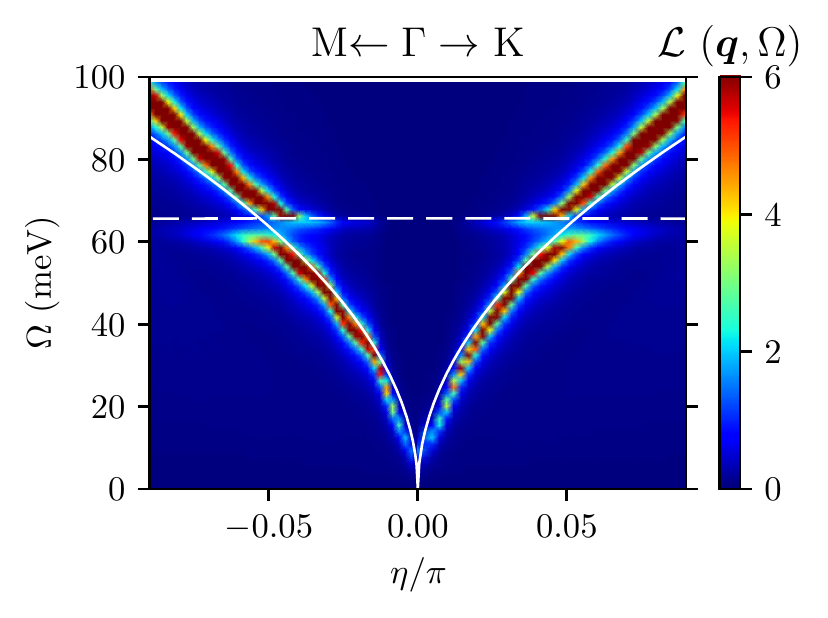}
\put(-1,26){\rm (b)}
\end{overpic} \\
\begin{overpic}[unit=1mm,width=0.48\columnwidth]{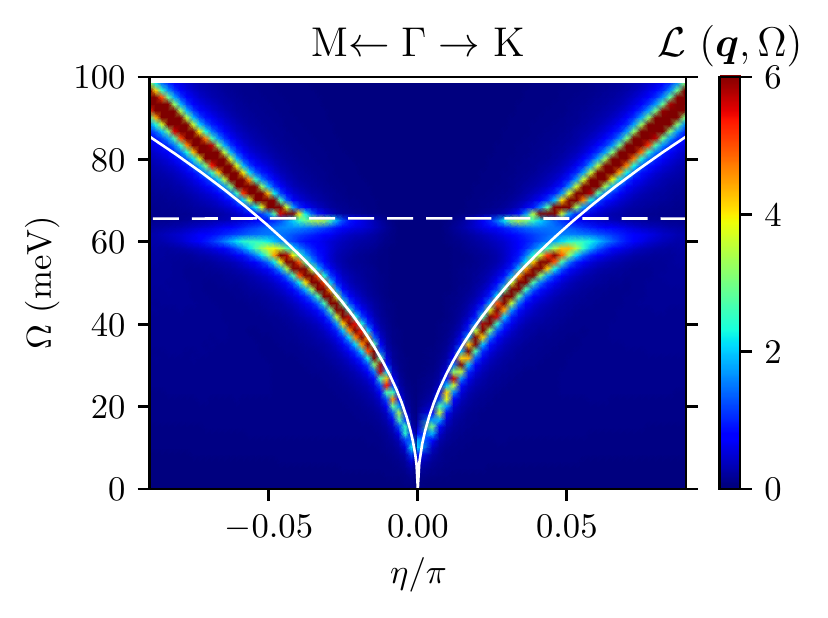}
\put(-1,26){\rm (c)}
\end{overpic} &
\begin{overpic}[unit=1mm,width=0.48\columnwidth]{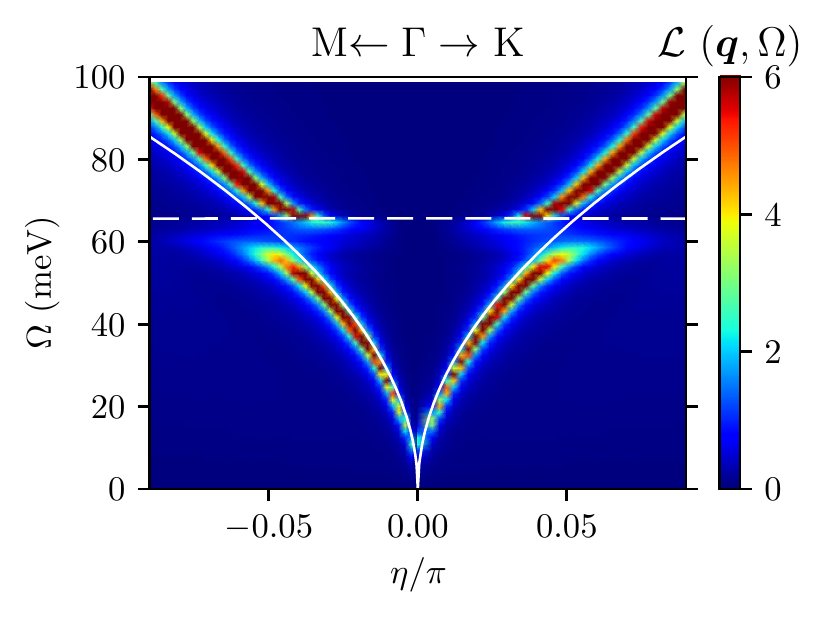}
\put(-1,26){\rm (d)}
\end{overpic} 
\end{tabular}
\caption{(Color online) The calculated electron energy-loss function ${\cal L}({\bm q}, \Omega)$ (color scale)---defined in Eq.~(11) of the main text---of monolayer doped Cr$_2$Ge$_2$Te$_6$ is plotted as a function of $\Omega$ and ${\bm q}$, around the $\Gamma$ point. The bare plasmon and bare magnon dispersions are shown by solid and dashed white lines, respectively. Results presented in this figure have been obtained by using the same parameters as in Fig.~1 of the main text, i.e.~$\kappa = 3.0$ and a $1 \%$ hole doping corresponding to a $\approx 2.5 \times 10^{12}~{\rm cm}^{-2}$ hole density. Different panels refer to different values of temperature: panel (a) $T= 10~{\rm K}$, panel (b) $T= 15~{\rm K}$, panel (c) $T= 20~{\rm K}$, and panel (d) $T= 25~{\rm K}$.\label{fig:lossTemp}}
\end{figure}

We evaluated $\gamma({\bm q},\Omega)$ and the electron energy-loss function ${\cal L}({\bm q}, \Omega)$ for $T\ll T_{\rm C}$ but it important to discuss the  $T$-dependence of $\gamma ({\bm q},\Omega)$ because it  affects the plasmon-magnon interaction. Owing to the presence of magnon bubble, $\gamma ({\bm q},\Omega)$ has a divergence near the upper magnon branch frequency, and the $\Omega$ and $T$-dependence of $\gamma({\bm q},\Omega)$ are determined by the magnon part. Moreover, as discussed before,  $\gamma ({\bm q},\Omega)$ has a negligible dependence on ${\bm q}$ in the long-wavelength limit. Further calculations reveal that the temperature dependence of $\gamma$ is controlled by the thermal factor $n_{\rm B} (\Omega_{0,-}^{\rm (m)}) - n_{\rm B} (\Omega_{0,+}^{\rm (m)})$, $\Omega_{0,\pm}^{\rm (m)}$ being the frequencies of the two magnon branches for ${\bm q} = {\bm 0}$. Here, we are interested in temperatures below $T_{\rm C}$ ($\approx 34~{\rm K}$), which is much smaller than the upper magnon branch energy ($\approx~760~{\rm K}$). Therefore, $n_{\rm B} (\Omega_{0,+}^{\rm (m)}) \ll n_{\rm B} (\Omega_{0,-}^{\rm (m)})$, and we have
\begin{equation}
\gamma ({\bm q},\Omega) \sim \dfrac{1}{e^{{\Omega_{0,-}^{\rm (m)}}/(k_{\rm B}T)}-1}~,
\label{Eq:gamma}
\end{equation}
where $\Omega_{0,-}^{\rm (m)} \approx 268~{\rm \mu eV}$ ($\approx 3.1~{\rm K}$). In the temperature range ($T \approx 10$-$30~{\rm K}$) we are interested in, we approximate
\begin{equation}
\gamma ({\bm q},\Omega) \sim \dfrac{k_{\rm B}T}{\Omega_{0,-}^{\rm (m)}}~.
\end{equation}
The temperature dependence of $\gamma ({\bm q},\Omega)$ is shown in Fig.~\ref{fig:gamma2}(d). Evidently, the linear dependence of $\gamma ({\bm q},\Omega)$ on $T$ is well described by Eq.~(\ref{Eq:gamma}). In summary, $\gamma({\bm q},\Omega)$ is small at low temperatures, where we expect that the plasmon-magnon interaction is suppressed. On the other contrary, the plasmon-magnon interaction increases with increasing $T$ because $\gamma({\bm q},\Omega)$ increases, quenching the plasmon oscillation. Numerical results for the $T$ dependence of the loss function confirming these expectations are reported in Fig.~\ref{fig:lossTemp}.

\begin{thebibliography}{99}
\bibitem{footnote_DM}
The nearest-neighbor Dzyaloshinski-Moriya interaction (DMI) term, $\hat{\cal H}_{\rm DMI} =\sum_{\langle i,j\rangle}{\bm D}_{ij}\cdot(\hat{\bm S}_i\times\hat{\bm S}_j)$, is absent in Eq.~(\ref{eq:Heisenberg+MAE}) due to the existence of an inversion center between the ${\rm Cr}$ ions. The second-nearest-neighbor DMI interaction term is in principle non-zero, but we have not considered it here as it does not affect spin waves in a significant manner, in the long-wavelength limit.
\end{thebibliography}
\end{document}